\documentclass[preprint,review,12pt]{elsarticle}
\usepackage{natbib}
\usepackage{amssymb}
\usepackage{amsmath}
 \usepackage{color}
\usepackage{graphicx}
\usepackage{dcolumn}
\usepackage{bm}
\usepackage{hyperref}
\usepackage{doi}
\usepackage{url}          
\usepackage{booktabs}      
\usepackage{nicefrac}       
\usepackage{microtype}     
\usepackage{physics}
\usepackage{orcidlink}
\usepackage{cuted}
\usepackage{threeparttable}
\usepackage[margin=1.375in]{geometry}
\usepackage{caption}

\DeclareCaptionFont{pt}{\fontsize{12pt}{12pt}\selectfont}
\journal{xxx}

\begin{document}

\begin{frontmatter}

\title{Learning in Wilson-Cowan model for metapopulation}

\author[1]{Raffaele Marino\corref{cor1}\orcidlink{0000-0002-2311-4380}}
\cortext[cor1]{Corresponding author}
\ead{raffaele.marino@unifi.it}
\author[1]{Lorenzo Buffoni} 
\author[1]{Lorenzo Chicchi} 
\author[2]{Francesca Di Patti} 
\author[1]{Diego Febbe}
\author[1]{Lorenzo Giambagli}
\author[1]{Duccio Fanelli}

\affiliation[1]{organization={Department of Physics and Astronomy, University of Florence},
            addressline={Via Giovanni Sansone  1}, 
            city={Sesto Fiorentino},
            postcode={50019}, 
            state={Florence},
            country={Italy}}

\affiliation[2]{organization={Department of Mathematics and Computer Science, University of Florence},
            addressline={ Viale Morgagni 67}, 
            city={Firenze},
            postcode={50134}, 
            state={Firenze},
            country={Italy}}

\begin{abstract}
The Wilson-Cowan model for metapopulation, a Neural Mass Network Model, treats different subcortical regions of the brain as connected nodes, with connections representing various types of structural, functional, or effective neuronal connectivity between these regions. Each region comprises interacting populations of excitatory and inhibitory cells, consistent with the standard Wilson-Cowan model. In this paper, we show how to incorporate stable attractors into such a metapopulation model’s dynamics. By doing so, we transform the Neural Mass Network Model into a biologically inspired learning algorithm capable of solving different classification tasks. We test it on MNIST and Fashion MNIST in combination with convolutional neural networks, as well as on CIFAR-10 and TF-FLOWERS, and in combination with a transformer architecture (BERT) on IMDB, consistently achieving high classification accuracy.
\end{abstract}


\end{frontmatter}

\section{Introduction}\label{sec:intro}

Understanding brain information processing requires building computational models that are capable of performing cognitive tasks \citep{kriegeskorte2018cognitive}. A brain computational model is a mathematical model that mimics the brain information processing underlying the performance of some task at some level of abstraction. At microscale, biological processes that underlie brain computation are described by biophysical models \citep{MURRAY2018777, pathak2022biophysical}, while, at macroscale, processes occurring in the brain are modeled by brain-dynamical and causal-interaction models \citep{karameh2019blind, sun2014discriminative}. In the past decades, the field of computational neuroscience has developed  many mathematical models of elementary computational components \citep{mcculloch1943logical, rosenblatt1958perceptron} and their implementation with biological neurons \citep{abbott2008theoretical}. In this paper, we discuss one of them: the Wilson-Cowan model \citep{wilson1972excitatory}.

The Wilson-Cowan model describes the evolution of excitatory and inhibitory activity in a synaptically coupled neuronal network. As opposed to being a detailed biophysical model, it is a coarse-grained description of the overall activity of a large-scale neuronal network, employing just two differential equations \citep{kilpatrick2022wilson}. As such, they embraced nonlinear dynamics, but in an interpretable form, i.e., they were motivated by physiological evidence \citep{mountcastle1997columnar},  which suggested the existence of certain populations of neurons with similar responses to external stimuli \citep{doi:10.1152/jn.1965.28.2.229}. Indeed, it was the first mathematical formulation to emphasize the significance of interactions between excitatory (E) and inhibitory (I) neural populations in cortical tissue \citep{Harms2013}, thereby incorporating both cooperation and competition \citep{wilson2021evolution}. 

This model has been widely used to study various aspects of neural activity: as a single-node description of excitatory-inhibitory population dynamics, as a building block for larger-scale brain network modeling studies, and as the underpinning of spatially extended models of neural dynamics at the tissue scale. It takes into account essential parameters such as the strength of synaptic connections among each type of neuronal population and the intensity of input received by each population. By manipulating these parameters, the model can replicate a range of dynamic brain behaviors, including multistability \citep{kaslik2024stability} and limit cycles \citep{wilson1972excitatory}. Other applications are: stable inhomogeneous steady states that store information dynamically and suggest a basis for short-term memory \citep{byrne2017learning}, oscillations \citep{maruyama2014analysis}, traveling waves \citep{harris2018traveling}, and the formation of spatial patterns \citep{wilson1973mathematical}. Additionally, the model captures information processing \citep{campbell1996synchronization}, binocular rivalry \citep{WILSON201789}, cognitive dynamics of movement \citep{erlhagen2002dynamic}, phase-amplitude coupling \citep{duchet2020phase}, neuroimaging data \citep{gonzalez2024fractional}, cortical resonant frequencies \citep{cowan2016wilson}, epilepsy \citep{wilson1973mathematical, meijer2015modeling}, and decision \citep{binder2009encyclopedia}, among other complex brain activities \citep{painchaud2022beyond, kora2023coarse}. Lastly, Wilson–Cowan models are a key component of the Virtual Brain project that aims to deliver the first simulation of the human brain based on personalized large-scale connectivity \citep{sanz2015mathematical}.

Building upon these foundational methods, the Wilson–Cowan model has been extended providing a deeper insight into the emergent collective behaviors of networks across multiple scales of organization \citep{lucilla2}.
Most obviously, it is possible to generalize to multiple excitatory and inhibitory populations reflective of particular cortical areas and functions. 
For examples, \citep{conti2019role} studied the dynamics of a network of Wilson–Cowan model  (a system of connected Wilson–Cowan oscillators). By observing that information transfer within each cortical area is not instantaneous, they consider a system of delay differential equations with two different kinds of discrete delay for exploring  a variety of larger networks, in order to determine how the network topology will influence time delayed Wilson–Cowan dynamics. They find that network structure can regularize
or deregularize the dynamics. In \citep{sanchez2024personalized}, the authors, instead, use personalized Alzheimer’s disease computational models built on whole-brain Wilson-Cowan oscillators to evaluate the direct impact of toxic protein deposition on neuronal activity.

Although the applications of the Wilson-Cowan model seem to cover many areas, a fundamental gap, as far as we know, remains: these extensions have not yet been applied to perform cognitive tasks, such as learning patterns in image recognition or general classification tasks.

\begin{figure*}
\centering
\includegraphics[scale=0.20]{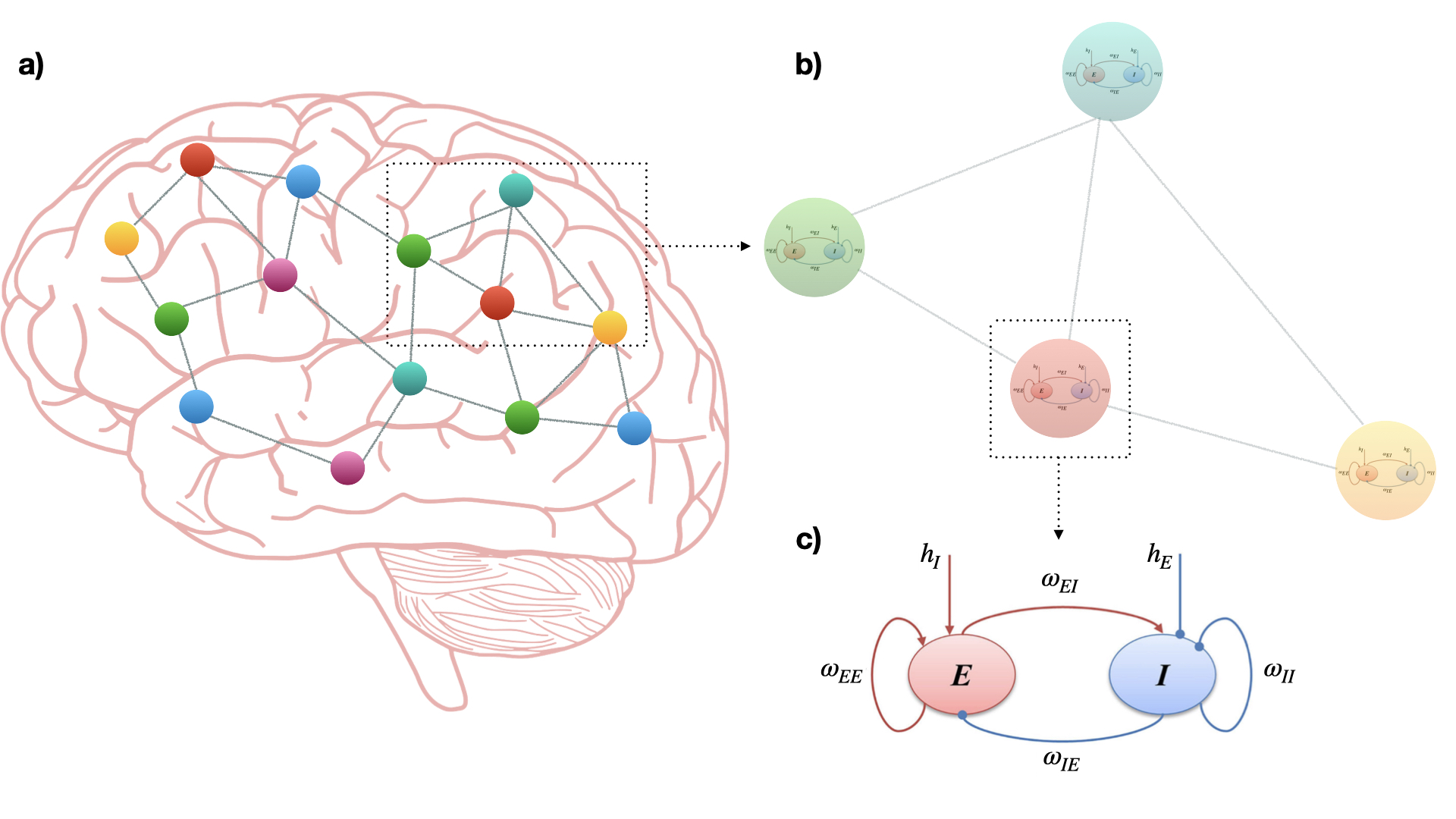}
\caption{\textbf{Cartoon of the  Neural Mass Network Model}. This figure illustrates a cartoon of the  Neural Mass Network Model. \textbf{a)} Nodes represent distinct brain regions. \textbf{b)} Within each region, or node of the network, there are interacting subpopulations of excitatory and inhibitory cells, in accord with the standard Wilson-Cowan model. \textbf{c)} Standard Wilson-Cowan model.}
\label{figmodel}
\end{figure*}

To fill this gap, in this paper, we consider a network metapopulation form of the Wilson-Cowan model (a Neural Mass Network Model \citep{byrne2020next}). Biologically speaking, such a network treats different subcortical regions of the brain as connected nodes, and the connections represent various types of links between them, as structural, functional, or effective connectivity between these distinct subcortical regions \citep{deweerdt2019map}. Within each region, or node of the network, there are interacting subpopulations of excitatory and inhibitory cells \citep{bazinet2023towards}, in accord with the standard Wilson-Cowan model (a cartoon  is displayed  in Fig. \ref{figmodel}). Moreover, noting that since long-range connections in the brain mainly project from excitatory pyramidal cells \citep{bekkers2011pyramidal, gerfen2018long}, we restrict couplings between brain regions to connections between only excitatory subpopulations. 

Given such a network, we aim to explore the learning capabilities and generalization abilities on classification tasks. To do so, we present, as in \citep{marino2023SANODE}, a method for planting stable attractors into the dynamics of this Neural Mass Network Model (as described in Sec. \ref{model::WClearning}, \ref{sec::linearstab}, \ref{sec::train} and illustrated in Fig. \ref{figtrain}). The presented method incorporates stable attractors into the model's dynamics, identifiable as targets for the classification task. These attractors are pairs of eigenvectors and eigenvalues of the adjacency matrix of the graph, selected a priori to meet the task requirements. Each eigenvector, used as long-term memory, consists of stable fixed points of the dynamics of each node. When the model reaches an attractor, the dynamics stabilizes, positioning on one of the targets of the classification task. 

Using machine learning techniques to learn the others eigenvectors and eigenvalues of the matrix, we validate the classification performance of such a model (and combinations with Convolutional Neural Networks (CNNs)) on MNIST dataset \citep{deng2012mnist} (grey scale handwritten images of size $28\times 28$), Fashion-MNIST dataset \citep{xiao2017fashion} (grey scale Zalando's article images of size $28\times 28$), CIFAR-10 dataset \citep{cifar} (colored images of size $32\times 32 \times 3$) and  TF-FLOWERS dataset \citep{Xu2022} (colored images of size $224\times 224\times 3$). Moreover, we tested our model in combination with a transformer architecture for classifying text reviews from the IMDB dataset \citep{zm1y-b270-20}.

The remaining sections of this paper are organized as follows.  Section \ref{model::WClearning} details our Wilson-Cowan model for metapopulation, showing also the possible biological connections of the computational model. Section \ref{sec::linearstab} presents
our methodology for enforcing stable attractors into the model. The methodology describing how to train our model for being a learning algorithm is in Section \ref{sec::train}, while the experimental numerical validations on well-known datasets are presented in Section \ref{sec::results}. Finally, in Section \ref{sec::conclusion}, we conclude the paper discussing and summarizing the obtained results.

\section{The model}\label{model::WClearning}

According to graph theory, brain networks can be described as graphs composed of nodes (vertices) representing neural elements (brain regions) linked by edges representing physical connections (synapses or axonal projections).

The topological and physical distances between elements in brain networks are intricate. Neurons and brain regions, containing populations of neurons, have a higher probability of being connected if they are spatially close, whereas connections between spatially remote neurons or brain regions are less likely \citep{bullmore2009complex}. This is because longer axonal projections are more expensive in terms of material and energy costs; indeed, the spatial layout of neurons or brain regions is economically arranged to minimize axonal volume \citep{bullmore2012economy}.

Given these biological observations, we propose that there are many interacting populations, forming a network metapopulation version of the Wilson-Cowan model. More precisely, we are given a graph $\mathcal{G}(\mathcal{N},\mathcal{E})$, where each of the $\mathcal{N}$ vertexes contains a population of neurons, composed by excitatory and inhibitory subpopulations, and interacts with another set of subpopulations contained into another vertex of the graph through a link in the set of $\mathcal{E}$. Each node in this model, therefore, corresponds to a population of neurons, including excitatory and inhibitory subpopulations, and hence we have a metapopulation model for some regions of the brain. Such populations can be finite or infinite. In this manuscript, we assume they are infinite. We will analyze the effect of finite size corrections at the single population level in future works.

The interactions between vertices are described, according to graph theory terminology, by an adjacency matrix $\mathbf{A}$ of the graph underlying the network. In general $\mathbf{A} \in \mathbb{R}^{\mathcal{N} \times \mathcal{N}}$. In this manuscript, an element of a matrix $\mathbf{A}$ is defined as $[\mathbf{A}]_{ln}$. The matrix $\mathbf{A}$, in our case, can be factorized by definition, i.e., $\mathbf{A}=\Phi  \Lambda \Phi^{-1}$, where $\Lambda$ is a diagonal matrix $ \Lambda \in \mathbb{R}^{\mathcal{N} \times \mathcal{N}}$, composed by the eigenvalues of $\mathbf{A}$, and $\Phi  \in \mathbb{R}^{\mathcal{N} \times \mathcal{N}}$ is a matrix where each column, i.e., $\vec{\phi}_{n}=[\Phi]_{:,n}$, with $n=1,\dots,\mathcal{N}$, is an eigenvector of $\mathbf{A}$. With the symbol $(\cdot)^{-1}$ we define the inverse matrix operation. The choice of dealing with the above decomposition of the coupling matrix echoes the spectral approach to Deep Learning discussed in \citep{chicchi2023complex} and reference therein. 

As stated before, each $i$-th vertex in $\mathcal{G}(\mathcal{N},\mathcal{E})$ has its own pair of Wilson-Cowan equations, describing the dynamics of a population of neurons, composed by subpopulations of excitatory and inhibitory species.
Each subpopulation can be described by a proportion of excitatory and inhibitory neurons firing per unit time at time $t$ \footnote{In this manuscript, all the variables are dimensionless.}. This proportion, in literature, is formalized with letter $x$ for excitatory neurons and letter $y$ for inhibitory neurons, respectively. Mathematically, the whole set of exitatory and inhibitory subpopulations can be described by the vectors $\vec{x}\in \mathbb{R}^{\mathcal{N}}$ and $\vec{y}\in \mathbb{R}^{\mathcal{N}}$. In this manuscript, we refer to the $i$-th component of a vector $\vec{v}$ with $[\vec{v}]_i$.  

For a sake of simplicity, we collect the two species of neurons into a matrix $\mathbf{z} \in \mathbb{R}^{\mathcal{N} \times 2}$. Mathematically, we obtain that $[\mathbf{z}]_{:,1}=\vec{x}$, $[\mathbf{z}]_{:,2}=\vec{y}$, $[\mathbf{z}]_{i,:}=([\vec{x}]_i,[\vec{y}]_i)$. In vector notation we have: $\vec{z}_{1}=[\mathbf{z}]_{:,1}$, $\vec{z}_{2}=[\mathbf{z}]_{:,2}$ and $\vec{\zeta}_{i}=([\mathbf{z}]_{i,:})^{\intercal}$, where $\vec{z}_{1}, \vec{z}_{2} \in \mathbb{R}^{\mathcal{N}}$, while $\vec{\zeta}_{i} \in \mathbb{R}^{2}$ and $^{\intercal}$ signifies the transpose operation.

Thus, the Wilson-Cowan equations of the model that govern the dynamics in each node of the network read:

\begin{subequations}\label{eq_stoc_vec}
\begin{align}
\frac{d[\vec{z}_{1}]_{i}(t)}{dt}&=-\alpha_E [\vec{z}_{1}]_{i}(t) +(1-[\vec{z}_{1}]_{i}(t))[\vec{f}_E(\vec{s_E}(\vec{z}_{1}(t), \vec{z}_{2}(t))]_i\\
\frac{d[\vec{z}_{2}]_{i}(t)}{dt}&=\frac{1}{\gamma}\left(-\alpha_I [\vec{z}_{2}]_{i}(t) +(1-[\vec{z}_{2}]_{i}(t))[\vec{f}_I(\vec{s_I}(\vec{z}_{1}(t),\vec{z}_{2}(t)))]_i\right)
\end{align}
\end{subequations}

where $i=1,\dots,\mathcal{N}$. Here, $0 \leq [\vec{z}_{1}]_{i}(t), [\vec{z}_{2}]_{i}(t) \leq 1$, $i=1,\dots, \mathcal{N}$, with $0$ corresponding to a state of quiescence in neuronal activity. 

The system also captures the refractory dynamics of both subpopulations, defined by the pre-factors $1\, – \,[\vec{z}_{1}]_{i}(t)$ and $1\, –\, [\vec{z}_{2}]_{i}(t)$, tracking the period of time during which cells are incapable of stimulation following an activation \footnote{This term has often been neglected in subsequent considerations of the model, and \citep{pinto1996quantitative} showed that it effectively rescales the parameters of the nonlinearities.}.  $\alpha_{E,I}$ are constant rate functions. 

The functions $\vec{f}_{E}(\cdot)$ and $\vec{f}_{I}(\cdot)$ represent nonlinear activation functions (non linear firing rate) for excitatory and inhibitory neurons, respectively, acting component-wise, i.e., $[\vec{f}_{E,I}]_{i}=f^{(1)}_{E,I}+f^{(2)}_{E,I} \tanh (\beta_{E,I} [\vec{s}_{E,I}]_i)$  where $\beta_{E,I}$ is the gain parameter. The quantities $f^{(1)}_{E,I}$ and  $f^{(2)}_{E,I}$ enter the definition of the sigmoidal non linear function and allow for a swift control of the location of fixed points. The offset $f^{(1)}_{E,I}$ sets in particular the degree of residual activity when $[\vec{s}_{E,I}]_{i}=0$. These nonlinearities capture the threshold-like behavior of neurons, i.e., a neuron only fires if its input exceeds a certain threshold. Obviously, they are functions of the currents  $\vec{s}_{E,I}$.

We define $\gamma$ as the ratio between inhibitors and activators in the population of neurons on a single vertex of the graph. This ratio is conventionally set to $0.25$. This is due to the fact that it is known that in the brain, there are  billions of neurons in the cortex. Among them, scientists have shown that $80\%$ are excitatory, whereas the remaining $20\%$ are inhibitory \citep{markram2004interneurons}, depending on whether they have a prominent postsynaptic density or a very thin postsynaptic density, respectively \citep{DEFELIPE1992563}.

We perform an important modification in the current vectors $\vec{s_E}(\vec{z}_{1}(t),\vec{z}_{2}(t))$ and $\vec{s_I}(\vec{z}_{1}(t),\vec{z}_{2}(t))$ with respect to the standard Wilson-Cowan model. Indeed, in these functions, we introduce the coupling term between the regions of the brain. 
More precisely, the equations for the currents are:

\begin{subequations}\label{OMSeieqs}
\begin{align}
    \vec{s_E}(\vec{z}_{1}(t),\vec{z}_{2}(t)) &= \omega_{EE} \vec{z}_{1}(t) - \omega_{EI} \vec{z}_{2}(t) + h_{E}\vec{e} +\Gamma \mathbf{A}\vec{z}_{1}(t)  \\
    \vec{s_I}(\vec{z}_{1}(t),\vec{z}_{2}(t)) &= \omega_{IE} \vec{z}_{1}(t) - \omega_{II} \vec{z}_{2}(t) + h_{I}\vec{e} 
\end{align}
\end{subequations}

where $\vec{e} \in \mathbb{R}^{\mathcal{N}}$  is a vector where all components are set to $1$, $\Gamma=1/\sqrt{\mathcal{N}}$ is a scaling factor \citep{ marino2023phase, huang2021statistical}, $\omega_{EE}$, $\omega_{II}$, $\omega_{EI}$, and $\omega_{IE}$ denote the strength of connectivity within and between the excitatory and inhibitory subpopulations. $h_{E}$ and $h_{I}$ denote the strength of the external input to each of these subpopulations. The $\Gamma \mathbf{A}\vec{z}_{1}(t)$ term couples the nodes of the network, and gives the strength of the interactions between connected nodes \citep{PIETRAS20191, PhysRevE.93.062147}. We recall that $\mathbf{A}\vec{z}_{1}(t)$ is a multiplication between a matrix in $\mathbb{R}^{\mathcal{N} \times \mathcal{N}}$ and a vector in $\mathbb{R}^{\mathcal{N}}$. It is worth pointing out that inter-node connections are established between excitatory subpopulations only, as is true of the brain. 
Such a term has been introduced in many works, as \citep{ Abeysuriya2018, Pinder2021},
for analyzing different dynamics in Wilson-Cowan oscillator networks, including approaches that use a Laplacian instead of an adjacency matrix. In all these cases, the authors analyze only the dynamical behaviors, without embedding stable attractors or applying the resultant dynamics to a learning task. As far as we know, our work is the first in this direction of research.

\begin{figure*}
\centering
\includegraphics[scale=0.5]{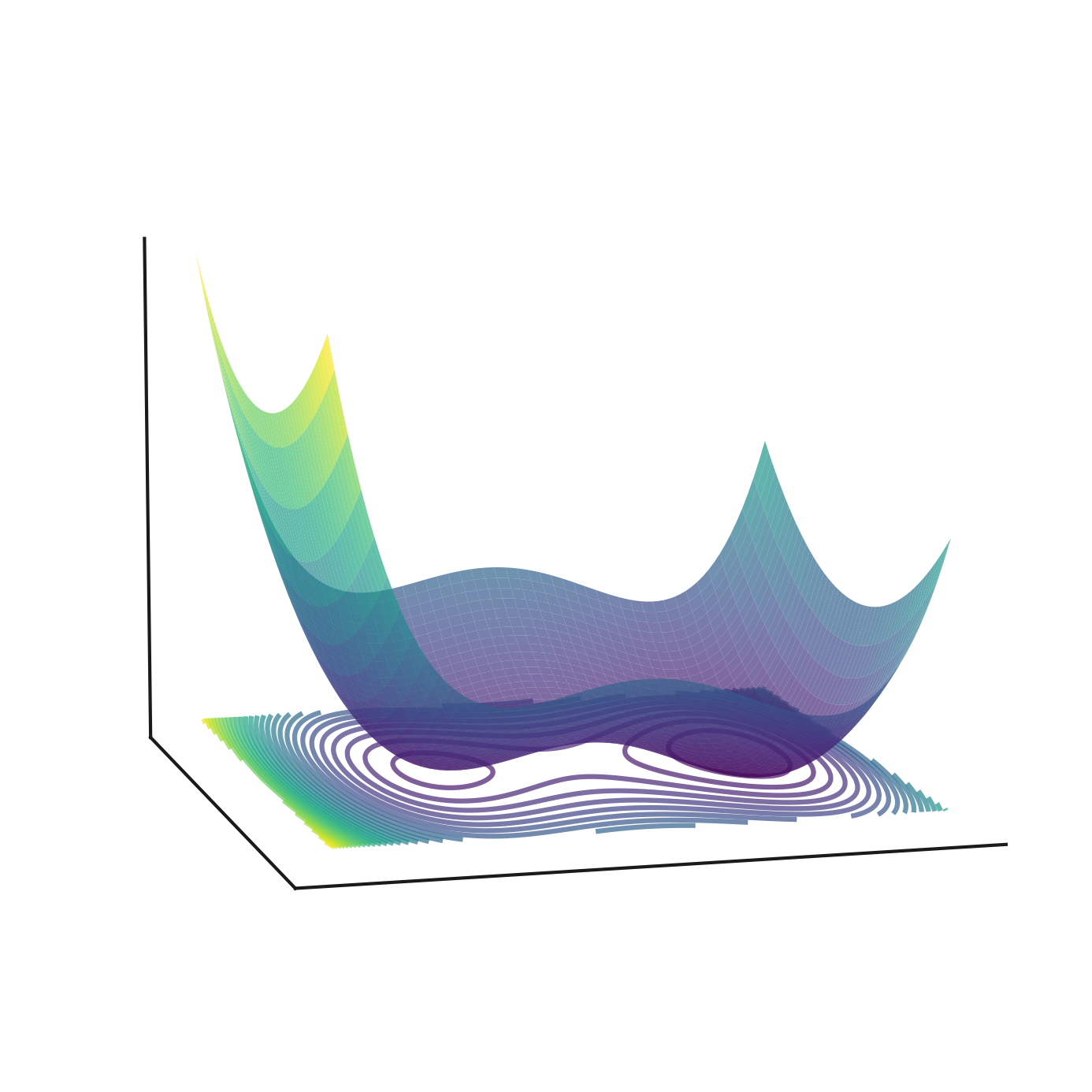}
\caption{\textbf{Cartoon of the  bistable profile}. This figure illustrates a cartoon of the bistable profile of the standard Wilson-Cowan model we adopt in each node of the network.}
\label{bistability}
\end{figure*}

The system's evolution can lead to unique fixed points, multiple fixed points each with its own basin of attraction, or even result in an oscillating periodic attractor, depending on the choice of parameters in equations \eqref{eq_stoc_vec} and \eqref{OMSeieqs}. Throughout this paper, our primary focus will be on parameter configurations leading to bistability \citep{golpayegan2023bistability} (Fig. \ref{bistability} displays a cartoon of the bistable profile for the standard Wilson-Cowan model). We select the bistabile profile to construct a dictionary for our classification task. Specifically, we aim to utilize the fixed points of the oscillator as targets for the dynamics to converge to.
We, therefore, select parameters in equation \eqref{eq_stoc_vec} and \eqref{OMSeieqs} to ensure the emergence of precisely two fixed points at every node of the graph (in the case the coupling term is omitted, i.e., when $\Gamma=0$), as used in \citep{zankoc2017diffusion}. We denote these solutions as $\vec{\overline{\zeta}}^{(1)}=(\overline{x}^{^{(1)}},\overline{y}^{^{(1)}})^{\intercal}$ and $\vec{\overline{\zeta}}^{(2)}=(\overline{x}^{^{(2)}},\overline{y}^{^{(2)}})^{\intercal}$ for each node. 

Our objective is to classify $K$ items based on the output of the network $\mathcal{G}(\mathcal{N}, \mathcal{E})$.
The network which defines the backbone of the examined dynamical model is made of $\mathcal{N}$ nodes, $\mathcal{N}$ denoting, for example, the number of pixels of the images to be classified. Each node is assigned with a standard Wilson-Cowan model. Nodes are then sensing its nearest neighbours, as stipulated by a linear coupling term which selectively acts on the excitatory species. The web of inter-nodes connections is stored in the  $\mathcal{N} \times \mathcal{N}$ adjacency weighted matrix ($\mathbf{A}=\Phi  \Lambda \Phi^{-1}$), whose elements represent the weights to be optimized in the training process. The above matrix is formulated in spectral domain \citep{marino2023SANODE}: a subset of $K$  suitably tailored eigenvectors is assigned to its kernel and define the attractive poles for the globally coupled dynamics. 
The eigenvalues populate a specific region of the real axis that yields stability of the crafted attractors, as dictated by the linear stability analysis presented in the next section. In essence the model to be trained is an extended collection of interacting units subject to contrasting tendencies: local reactions occur between subpopulations referred to a single brain region (or node), while global, spatially (across nodes) extended, interactions make the system to evolve towards complex heterogeneous stable states. In summary, the item to be classified is supplied as an initial condition of the dynamical model and the ensuing stable attractor, as reached by the trained model, flags for the corresponding classification label. Under this perspective, learning amounts to shaping the basin of attraction of the underlying dynamical model.

Having in mind that, we encode $K$ attractors within the matrix $\Phi$, representing distinct system configurations that it may evolve towards over time.  We specify $K$ columns of $\Phi$, thus $\vec{\phi}_{k}$ for $k=1,\dots,K$. By design, these column vectors are chosen permutations of the fixed points identified in the dynamics described by equation \eqref{eq_stoc_vec}, such that each element of $\vec{\phi}_{k}$ corresponds to either $[\vec{\overline{\zeta}}^{(1)}]_{1}$ or $[\vec{\overline{\zeta}}^{(2)}]_{1}$, namely $[\vec{\phi}_{k}]_i=$ $\{[\vec{\overline{\zeta}}^{(1)}]_{1}$ or $[\vec{\overline{\zeta}}^{(2)}]_{1}\}$ $\forall i=1,\dots,\mathcal{N}$ (see Fig. \ref{figtrain}, Panel \textbf{b)} and \textbf{c)}). Furthermore, we align only the eigenvectors $\vec{\phi}_{k}$, $k=1,\dots, K$, within the kernel of $\mathbf{A}$, highlighting that these vectors, as elements of the kernel, can be constructed from linear combinations of the kernel's orthonormal vectors. 

In conclusion, $\mathbf{A} \vec{\phi}_{k}=\lambda_{k}\vec{\phi}_{k}=\vec{0}$. Thus, each eigenvalue $\lambda_k$ associated to a $\vec{\phi}_{k}$ ($k=1,\dots,K$) in $\Lambda$ is set to zero. To streamline our approach without sacrificing generality, we choose the system parameters so that $[\vec{\overline{\zeta}}^{(1)}]_{2}=[\vec{\overline{\zeta}}^{(2)}]_{2}$. Although $[\vec{\overline{\zeta}}^{(1)}]_{2}$ represents a degree of freedom within the system dynamics, it lacks informational value for our classification task. Consequently, we refrain from adding an eigenvector composed exclusively of $[\vec{\overline{\zeta}}^{(2)}]_{2}$ values to $\mathbf{A}$. 

The embedded eigenvectors must be stable. To ensure this, we perform a linear stability analysis \citep{strogatz2018nonlinear} to identify the conditions that the non-zero eigenvalues in $\Lambda$ must satisfy in order to make $\vec{\phi}_{k}$ ($k=1,\dots,K$) stable attractors.

\section{Enforcing Linear Stability} \label{sec::linearstab}

Our aim is to enforce stability on the $K$ planted attractors in $\mathbf{A}$. To be stable a stationary state must be such that under a small perturbation, the dynamics of the system returns on it. For this reason, we consider the system of real first order differential equations $d\mathbf{z}/dt = G(\mathbf{z})$, where $G:\mathbb{R}^{\mathcal{N} \times 2}\to \mathbb{R}^{\mathcal{N} \times 2}$  identifies the functional components in equation \eqref{eq_stoc_vec}, thus the $i$-th row of $G(\mathbf{z})$ is, in matrix notation $d\vec{\zeta}_i/dt = ([G(\mathbf{z})]_{i,:})^{\intercal}$, equal to equation \eqref{eq_stoc_vec}.
At the steady state $d\vec{\zeta}_i/dt=0$, for all $i=1, \dots, \mathcal{N}$. Hence, $G(\mathbf{z}^*)=\mathbf{0}$.

For analyzing the stability of fixed points in the stationary state, i.e., $\mathbf{z}^*$,  we examine the behaviour of orbits near it.

Thus, we assume that $\mathbf{z}(t)=\mathbf{z}^*+\delta\mathbf{z}(t)$, where $\delta \mathbf{z}(t)$ is a small perturbation close to $\mathbf{z}^*$. Substituting into $d \mathbf{z}/dt = G(\mathbf{z})$, we can expand $G(\mathbf{z})$ to first order in $\delta\mathbf{z}(t)$, obtaining:
\begin{equation}
    G(\mathbf{z}^*+\delta\mathbf{z}(t))=G(\mathbf{z}^*)+\mathbf{J}(\mathbf{z}^*) \cdot \delta{\mathbf{z}}(t)+O(\delta\mathbf{z}^2(t)).
\end{equation}

The tensor $\mathbf{J}(\mathbf{z}^*)$, which denotes the Jacobian, can be easily computed. Indeed, by definition of Jacobian:
\begin{equation}
\mathbf{J}(\mathbf{z})=
\begin{bmatrix}
    \frac{d[G(\mathbf{z})]^{\intercal}_{1,:}}{d\vec{\zeta}_1} & \frac{d[G(\mathbf{z})]^{\intercal}_{1,:}}{d\vec{\zeta}_2} & \dots & \frac{d[G(\mathbf{z})]^{\intercal}_{1,:}}{d\vec{\zeta}_{\mathcal{N}}}\\
    \frac{d[G(\mathbf{z})]^{\intercal}_{2,:}}{d\vec{\zeta}_1} & \frac{d[G(\mathbf{z})]^{\intercal}_{2,:}}{d\vec{\zeta}_2} & \dots & \frac{d[G(\mathbf{z})]^{\intercal}_{2,:}}{d\vec{\zeta}_{\mathcal{N}}}\\
    \vdots & \vdots & \vdots & \vdots\\
    \frac{d[G(\mathbf{z})]^{\intercal}_{\mathcal{N},:}}{d\vec{\zeta}_1} & \frac{d[G(\mathbf{z})]^{\intercal}_{\mathcal{N},:}}{d\vec{\zeta}_2} & \dots & \frac{d[G(\mathbf{z})]^{\intercal}_{\mathcal{N},:}}{d\vec{\zeta}_{\mathcal{N}}}\\
\end{bmatrix}.
\end{equation}

Making the whole calculation together, we obtain the following equation for the time dependence of the perturbation of $\mathbf{z}$ from the steady state:
\begin{equation}\label{eq_stab1}
    \frac{d \delta \mathbf{z}(t)}{dt}=\mathbf{J}( \mathbf{z}^*) \cdot \delta \mathbf{z} + O(\delta \mathbf{z}^2).
\end{equation}
The linearized stability problem is obtained by neglecting terms of second order in $\delta \mathbf{z}$.

The Jacobian is in $\mathbb{R}^{(\mathcal{N} \times 2) \times (\mathcal{N}\times 2)}$. From \eqref{eq_stab1}, we can make it diagonal. In such a situation, we can compute only  a single component of the diagonal for computing the whole Jacobian. Mathematically, each component of the diagonalized Jacobian is a matrix $2 \times 2$ of components:
\begin{equation}
    [\mathbf{J}]_{ii}(\mathbf{z})= 
\begin{bmatrix}
    J_{[\vec{\zeta}_i]_1 [\vec{\zeta}_i]_1} & J_{[\vec{\zeta}_i]_1  [\vec{\zeta}_i]_2} \\
    J_{[\vec{\zeta}_i]_2 [\vec{\zeta}_i]_1} & J_{[\vec{\zeta}_i]_2  [\vec{\zeta}_i]_2} 
\end{bmatrix},
\end{equation}

where $[[\mathbf{J}]_{ii}(\mathbf{z})]_{el}$, with $e,l=1,2$, defines the derivatives with respect to species $e$ on the expression of the species $l$ on the dynamics of our model \eqref{eq_stoc_vec} on vertex $i$.

To solve \eqref{eq_stab1} (neglecting $O(\delta\mathbf{z}^2)$), we solve $ d \delta\vec{\zeta}_i(t)/dt=[\mathbf{J}]_{ii}(\mathbf{z}^*) \cdot \delta\vec{\zeta}_i$ on a single vertex $i$ of the network. We recall that the last equation has been obtained by observing that $\delta \vec{\zeta}_i=[\delta \mathbf{z}]_{i,:}$. Thus, by seeking a solution of the form $\delta\vec{\zeta}_i(t)=\vec{\chi}_i\exp(st)$ with $\vec{\chi}_i \in \mathbb{R}^2$, the problem becomes a simple eigenvalue problem: $[\mathbf{J}]_{ii}(\mathbf{z}^*) \vec{\chi}_i= s\vec{\chi}_i$  which has nontrivial solutions for values of $s$ satisfying the second order polynomial equation $D(s)=\text{det}([\mathbf{J}]_{ii}(\mathbf{z}^*)-s\mathbf{I})=0$, where $\mathbf{I}$ denotes a $2 \times 2$ identity matrix.

To look for asymptotic stability,  it suffices to consider two roots of $D(s)=0$, i.e. $s^{(+)}$ and $s^{(-)}$. At each root is associated an eigenvector $\vec{\chi}^{(\pm)}_i$. Any time evolution  can be represented as $\delta\vec{\zeta}_i(t)= c_{+}\vec{\chi}^{(+)}_i \exp(s^{(+)}t) + c_{-}\vec{\chi}^{(-)}_i \exp(s^{(-)}t)$, where $c_{\pm}$ are constant coefficients that can be  determined by the initial condition $\delta\vec{\zeta}_i(0)=c_{+}\vec{\chi}^{(+)}_i + c_{-}\vec{\chi}^{(-)}_i$. 

Given that the coefficients of the characteristic polynomial $D(s)$ are real, we expect that the eigenvalues $s^{(\pm)}$ to be real or appear in complex conjugate pairs.

By definition of stability, the real part of the eigenvalues $s^{(\pm)}$ must be negative. This implies that when  solved the following expression:

\begin{equation}\label{eqsecordersk}
s^{(\pm)}= \frac{\text{Tr}([\mathbf{J}]_{ii}(\mathbf{z}^*)) \pm \sqrt{\text{Tr}([\mathbf{J}]_{ii}(\mathbf{z}^*))^2 - 4 \text{det}([\mathbf{J}]_{ii}(\mathbf{z}^*))}}{2},
\end{equation}

the maximum value between the two real part of \eqref{eqsecordersk}, i.e., $\max\{\text{Re}(s^{(+)}), \text{Re}(s^{(-)})\}$, must be taken.
If such a value is negative, then the associated fixed point is asymptotically stable \citep{strogatz2018nonlinear}.

\begin{figure*}
\centering
\includegraphics[scale=0.5]{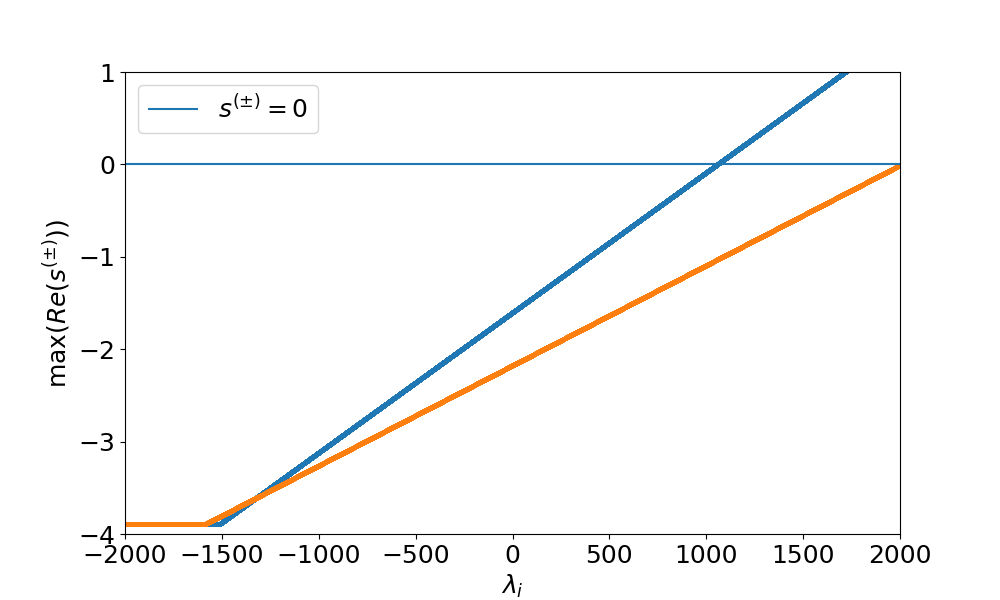}
\caption{\textbf{Stability before training.} The figure shows the maximum value of the real part of the eigenvalues $s^{(\pm)}$ as a function of the parameter $\lambda_i$. The two curves, with different colors, identify respectively the two different steady states. 
The parameters here employed read $\omega_{II}=1$, $\omega_{IE}=0.0$, $\omega_{EI}=2$, $\omega_{EE}=7.2$
$\alpha_E=1.5$, $\alpha_I=0.4$, $h_E=-1.2$, $h_I=0.1$, $\gamma=0.25$, $\beta_E=3.7$, $f^{(1)}_{E}=0.25$, $f^{(2)}_{E}=0.65$, $\beta_I=1$,
$f^{(1)}_{I}=0.5$, $f^{(2)}_{I}=0.5$, $\mathcal{N}=784$.}
\label{fig_stability}
\end{figure*}

To be explicit, the elements of $[\mathbf{J}]_{ii}(\vec{\mathbf{z}}^*)$ have the following expressions:

\begin{subequations}
\begin{align}
J_{[\vec{\zeta}_i]_1 [\vec{\zeta}_i]_1} &=-\alpha_E - [\vec{f_E}(\vec{s_E}([\mathbf{z}^*]_{:,1}, [\mathbf{z}^*]_{:,2}))]_i + (1-[\vec{\zeta}_i]_1)[\vec{f^{'}_E}(\vec{s_E}([\mathbf{z}^*]_{:,1}, [\mathbf{z}^*]_{:,2}))]_i(\omega_{EE}+\Gamma \lambda_i)\\
J_{[\vec{\zeta}_i]_1 [\vec{\zeta}_i]_2}&=(1-[\vec{\zeta}_i]_1)[\vec{f^{'}_E}(\vec{s_E}([\mathbf{z}^*]_{:,1}, [\mathbf{z}^*]_{:,2}))]_i(- \omega_{EI}) \\
J_{[\vec{\zeta}_i]_2 [\vec{\zeta}_i]_1}&= \frac{1}{\gamma}\left((1-[\vec{\zeta}_i]_2)[\vec{f^{'}_I}(\vec{s_I}([\mathbf{z}^*]_{:,1}, [\mathbf{z}^*]_{:,2}))]_i(\omega_{IE})\right)\\
J_{[\vec{\zeta}_i]_2 [\vec{\zeta}_i]_2}&= \frac{1}{\gamma}\left(-\alpha_I -[\vec{f_I}(\vec{s_I}([\mathbf{z}^*]_{:,1}, [\mathbf{z}^*]_{:,2}))]_i +(1-[\vec{\zeta}_i]_2)[\vec{f^{'}_I}(\vec{s_I}([\mathbf{z}^*]_{:,1}, [\mathbf{z}^*]_{:,2}))]_i(- \omega_{II})\right),
\end{align}
\end{subequations}

where $\vec{f^{'}_{E/I}}$ are the derivatives with respect to the species ${E/I}$.
$\lambda_i$ is the eigenvalue of the matrix $\mathbf{A}$, associated to the $i-$th vertex. The eigenvalue problem, the one associated to the stability problem, now contains the parameter $\lambda_i$, and equation \eqref{eqsecordersk} becomes parametric.  In such a situation, we can compute the entire region  of stability, and know exactly what kind of values the parameter $\lambda_i$ must take so that the stationary points are stable. In Fig. \ref{fig_stability}, we show the maximum value of the real part of the eigenvalues $s^{(\pm)}$ as a function of $\lambda_i$, on a single vertex $i$. In this case, the plot shows that the stationary points are stable as soon as the maximum value of the real part of the eigenvalues $s^{(\pm)}$ is negative. Through this method, we have identified the regions of $\lambda_i$ that ensure the stability of our stationary state, thereby securing our planted attractors.

\section{Training}\label{sec::train}

Obtained the stability criteria for our attractors, we can construct a statistical learning model for our classification task. In such a statistical learning model only the components of the matrix $\Phi$, which are not the embedded eigenvectors, the non-zero eigenvalues in $\Lambda$, and the $\gamma$ parameter can be learned. In other words, the matrix $\Phi$ should be regarded as a matrix where the first $K$ columns, identifying the embedded attractors of the system's dynamics, are not to be learned, just as the first $K$ eigenvalues of the matrix  $\Lambda$, which are fixed to be zero by construction. This approach ensures that the planted attractors guide the dynamics, letting the system to converge on selected attractors once the training has been completed. As an initial condition for the non-embedded eigenvectors, we opt for orthogonal random eigenvectors \citep{bansal2018can}. As an initial condition for $\gamma$ we opt for a value equal to $0.25$ (as described in Section \ref{model::WClearning}), while for the trainable eigenvalues we opt to initialize them with a normal distribution with mean equal to $-\sqrt{\mathcal{N}}$ and variance equal to $1$. The condition on the trainable eigenvalues satisfies completely the initial stability condition (see Fig. \ref{fig_stability}). Without loss of generality, we fix the non trainable parameters of the model, as described in the caption of Fig. \ref{fig_stability} and used in \citep{zankoc2017diffusion}.

Moving onto the model training, the parameter space for optimization resides in $ \mathbb{R}^{(\mathcal{N}-K)(\mathcal{N}+1)+ 1} $, where $ K $, we recall, denotes the number of classes in our classification problem, and the singular dimension relates to the parameter $\gamma$. 

In practice, we are in a supervised learning setting. Therefore, we have a dataset $ \mathcal{D} = (\mathbf{z}, \vec{\mathcal{T}})^{(j) \in [1, \dots, |\mathcal{D}|]} $ of size $ |\mathcal{D}| $, where $\mathbf{z}^{(j)}$ is an input datum, and $\vec{\mathcal{T}}^{(j)} $ represents the associated target.  For simplicity, the input datum is understood to be $\mathbf{z}^{(j)} \in \mathbb{R}^{\mathcal{N} \times 2}$. $K$ distinct and mutually different targets are mapped as attractors  of the dynamics. In other words, the target associated to the first class of items is the first eigenvector $\vec{\phi}_{1}$ of matrix $\mathbf{A}$, the target associated to the second class of items is the second eigenvector $\vec{\phi}_{2}$, and so on. We recall that the first $K$ eigenvectors of $\mathbf{A}$ have been planted with a particular structure, as described in Section \ref{model::WClearning}.

\begin{figure*}
\centering
\includegraphics[scale=0.25]{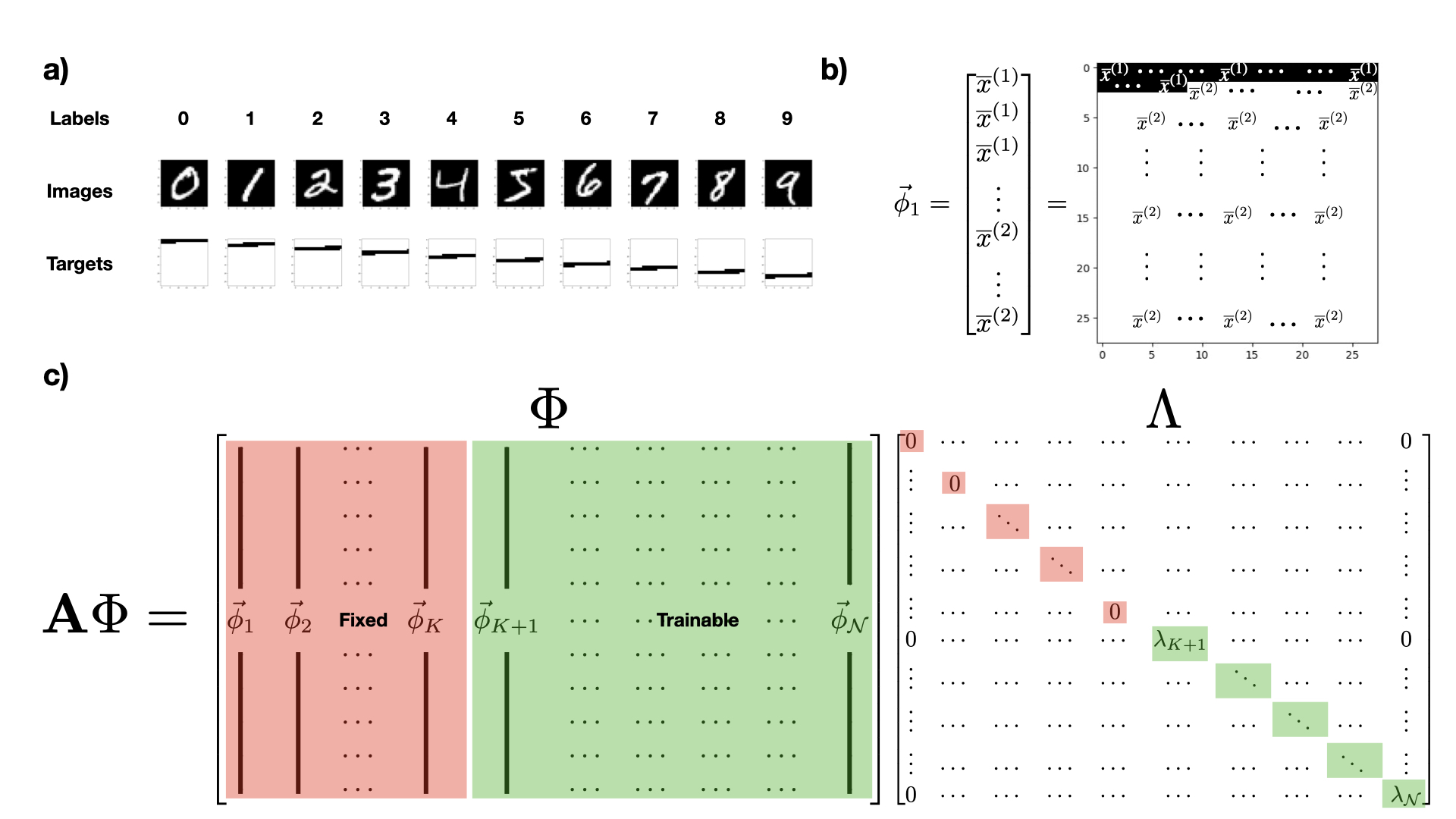}
\caption{\textbf{Building the model.} The figure illustrates the composition of the dataset and the method for introducing stable attractors into the dynamics. Panel \textbf{a)} displays the labels, images, and targets used in our model. The first row shows the labels of the MNIST image ($\mathcal{N}=784$) examples found in the second row. The third row presents the ten target images we created to serve as attractors for the dynamics. Panel \textbf{b)} shows the creation of an eigenvector of matrix $\mathbf{A}$. The image is composed solely of the excitatory stationary parts of the two fixed points in the Wilson-Cowan model. Each black pixel corresponds to a $\overline{x}^{(1)}$, while each white pixel corresponds to a $\overline{x}^{(2)}$. In general, to create $K$ target vectors for a dataset, we generate $K$ vectors of size $\mathcal{N}$, where $\frac{\mathcal{N}}{K+2}$ components are set to $\overline{x}^{(1)}$ and the remaining components to $\overline{x}^{(2)}$. Each unique target vector is constructed by setting the components from $k\frac{\mathcal{N}}{K+2}$ to $((k+1)\frac{\mathcal{N}}{K+2})-1$ to $\overline{x}^{(1)}$, with all other components set to $\overline{x}^{(2)}$, with $k=0,\dots,K-1$ that identifies the label of a single class.
Panel \textbf{c)} illustrates the positions of the fixed eigenvectors and eigenvalues (in red) and the trainable ones (in green). Recall that the matrix $\mathbf{A}$ is diagonalizable by definition.}
\label{figtrain}
\end{figure*}

For example, the MNIST dataset is composed of a set of $70000$ handwritten digit images of size  $28 \times 28$, ranging from $0$ to $9$, with associated labels for each digit. A label is just a number from $0$ to $9$, thus indicating the number of classes for the classification task is $K=10$.  By denoting the $j$-th  image of such dataset as $\vec{r}^{(j)} \in \mathbb{R}^{784}$, and its associated label with $\mathcal{T'}^{(j)} $, the $j$-th input element (our initial condition) of our dataset $ \mathcal{D}$ will be built as $\mathbf{z}^{(j)}=(\vec{z}^{(j)}_{1}=\vec{r}^{(j)},\vec{z}^{(j)}_{2}=\vec{r}^{(j)})$, with the associated target given by $\vec{\mathcal{T}}^{(j)}=\vec{\phi}_{\mathcal{T'}^{(j)}}$. Therefore, the $j$-th element of our dataset will be $(\mathbf{z}, \vec{\mathcal{T}})^{(j)}$, where $\mathbf{z} \in \mathbb{R}^{\mathcal{N} \times 2}$, $\vec{\mathcal{T}} \in \mathbb{R}^{\mathcal{N}}$, and $\mathcal{N}=784$. The cardinality of $\mathcal{D}$ will be $|\mathcal{D}|=70000$. This dataset is then split into a training and a test  set of size $|\mathcal{D}_{train}|=60000$ and $|\mathcal{D}_{test}|=10000$, respectively. Each node of the graph is associated with a pixel of an image that must be classified. The image evolves over time and reaches an asymptotically stable state corresponding to the one we initially planted. A figurative example of how the dataset is constructed and how stable attractors are planted is presented in Fig. \ref{figtrain}.

By construction, the data are sampled i.i.d. from  their unknown joint distribution $ \mathbf{P}(\mathbf{z},\vec{\mathcal{T}}) $. 

The objective of the training is to estimate the values of the model's weights in such a way that it is able to achieve its goal. This estimations is reached by the minimization of a loss function.
We define the loss function as:
\begin{equation}\label{eq::lossmatrix}
\mathcal{L}=\frac{1}{|\mathcal{D}| }\sum_{j=1}^{|\mathcal{D}| } (\vec{z}^{(j)}_{1}(T)-\vec{\mathcal{T}}^{(j)})^{\intercal}(\vec{z}^{(j)}_{1}(T)-\vec{\mathcal{T}}^{(j)}),   
\end{equation} 
where $T$ is a sufficiently large time, ensuring that a stationary state of the dynamical system is reached. 
The reader should note that in \eqref{eq::lossmatrix}, only the excitatory species is explicitly written. However, the inhibitory species is implicitly factored into the loss function through its contribution to the excitatory term, i.e., equation \eqref{eq_stoc_vec}.

The input datum $\mathbf{z}^{(j)}$ represents, for our dynamical system defined in \eqref{eq_stoc_vec},  an initial condition of the dynamics, i.e., $\mathbf{z}^{(j)}(t=0)$. From such initial condition, we evolve the dynamical system for a sufficiently large time $T$. The evolution is performed using Euler's algorithm, discretizing time into steps of $\Delta t=0.1$, unless explicitly specified otherwise. At each iteration, denoted with the index $n$, the system's state $\mathbf{z}^{(j)}_n$ is updated according to the equation: $\mathbf{z}^{(j)}_{n+1} =  \mathbf{z}^{(j)}_n + G(\mathbf{z}^{(j)}_n) \Delta t$, where $G(\mathbf{z}^{(j)})$ is the system's flux, defined in \eqref{eq_stoc_vec}. The total number of iterations, $n_{max}$, is determined a priori based on the desired simulation time $T$ and the chosen time step $\Delta t$ through the relation $n_{max} = \lfloor\frac{T}{\Delta t}\rfloor.$ This approach allowed us to efficiently simulate the system's long-term behavior by iteratively applying Euler's update rule. Once the system reaches the final iteration, we use the value of the solution to optimize the loss function.

To minimize the loss function in \eqref{eq::lossmatrix}, we specifically employ Accelerated Stochastic Gradient Descent using the Adam optimizer \citep{kingma2014adam}, with a learning rate of $0.1$, unless explicitly specified otherwise. The learning rate, the number of epochs needed to achieve optimal minimization of the loss function, the mini-batch size used in the Adam algorithm, and the time $T$ required to reach the stationary state have been optimized empirically (see \ref{App1}).

\begin{figure*}
\centering
\includegraphics[scale=0.25]{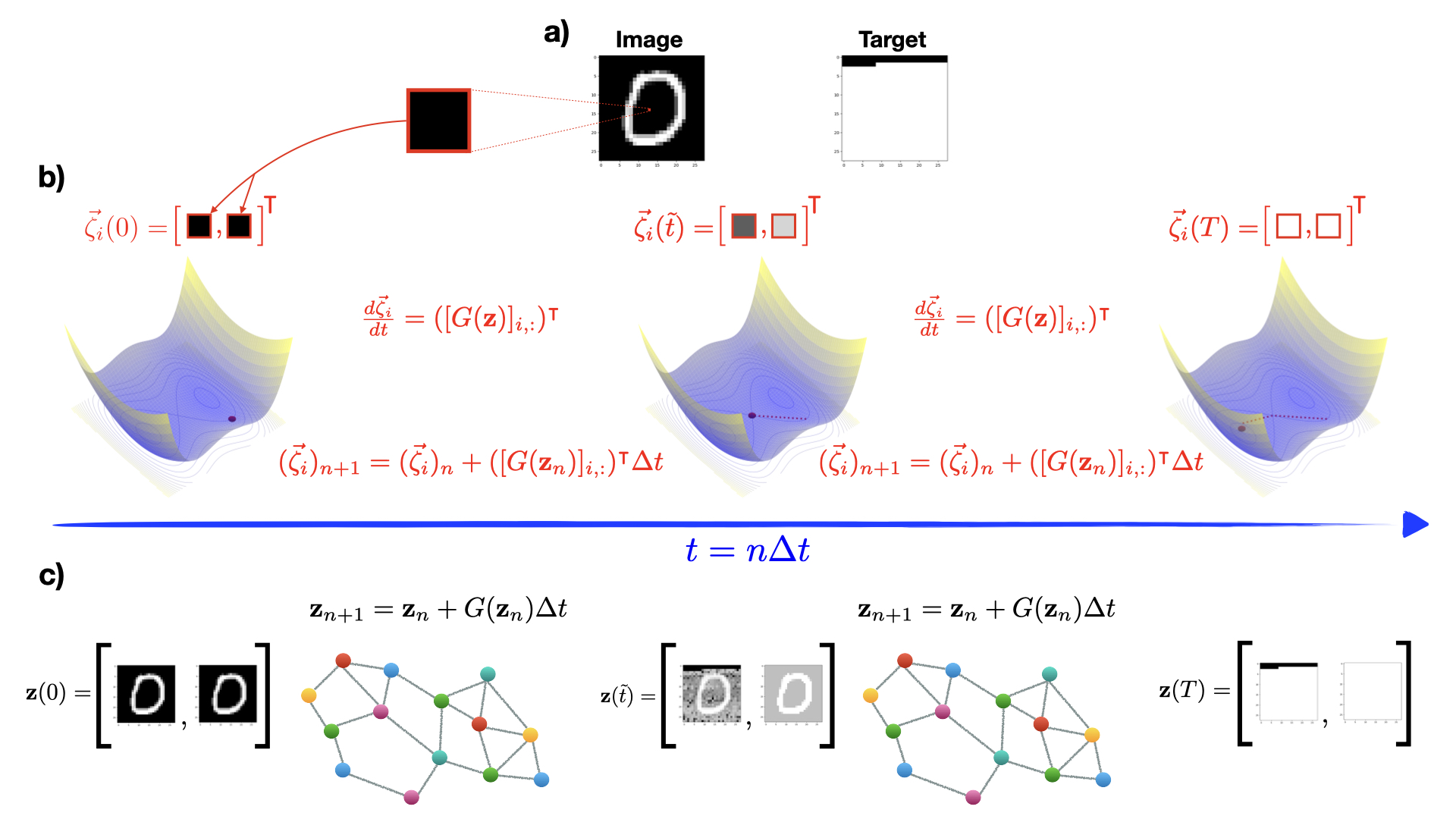}
\caption{\textbf{Dynamics of the model.} The figure illustrates a schematic representation of our Neural Mass Network Model. Panel \textbf{a)} displays an image from the MNIST ($\mathcal{N}=784$) dataset along with its associated target. The image features a black pixel with a red border, which we track throughout its dynamics. Panel \textbf{b)} shows this dynamic process. The equations are highlighted in red to emphasize that we are focusing on that single pixel. This pixel is depicted as a red sphere moving within a double well potential. Our model, we recall, is configured to exhibit a bistable profile. Initially, the vector $\vec{\zeta}_i$ is set to represent the black pixel for both subpopulations of neurons. As the dynamics evolve according to our model's equations, the pixel explores the bistable profile. It does not stop at the closest minimum; instead, it reaches the correct minimum designated for classification, thanks to the learned coupling in the matrix $\mathbf{A}$. Panel \textbf{c)} illustrates the overall dynamics of the entire image. Starting from an initial condition—the image we want to classify—the system uses the equations in \eqref{eq_stoc_vec} to achieve the final state, which is our stable attractor. The precision cutoff for this analysis has been set to the seventh decimal digit. }
\label{cartoondynamics}
\end{figure*}

Once the training procedure terminates, we have a classifier governed by a metapopulation of excitatory and inhibitory subpopulations. A sketch of the dynamics is presented in Fig. \ref{cartoondynamics}.

Quantifying the effectiveness of the Wilson-Cowan model for metapopulation as a learning algorithm  requires a well-defined performance metric. We choose to directly compare the generated output images with their corresponding target counterparts.

To do this, we compare the output image with all $K$  distinct and mutually different targets. This comparison yields a vector with $K$ components, where each component represents the normalized $L^2$ distance between the output image and the $k$-th target image, with $k = 1, \dots, K$. 
Mathematically, we have:
\begin{equation}
\vec{m}^{(j)}_f = \begin{pmatrix}
\frac{\sum_{i=1}^{\mathcal{N}}  [ (\vec{z}^{(j)}_1(T) - \vec{\mathcal{T}}^{(1)}) ]^2_i}{\sqrt{\sum_{i=1}^{\mathcal{N}}[ (\vec{z}^{(j)}_1(T)]^2_i \sum_{i=1}^{\mathcal{N}}[\vec{\mathcal{T}}^{(1)}) ]^2_i}}  \\
\frac{\sum_{i=1}^{\mathcal{N}}  [ (\vec{z}^{(j)}_1(T) - \vec{\mathcal{T}}^{(2)}) ]^2_i}{\sqrt{\sum_{i=1}^{\mathcal{N}}[ (\vec{z}^{(j)}_1(T)]^2_i \sum_{i=1}^{\mathcal{N}}[\vec{\mathcal{T}}^{(2)}) ]^2_i}}  \\
\vdots \\
\frac{\sum_{i=1}^{\mathcal{N}}  [ (\vec{z}^{(j)}_1(T) - \vec{\mathcal{T}}^{(K)}) ]^2_i}{\sqrt{\sum_{i=1}^{\mathcal{N}}[ (\vec{z}^{(j)}_1(T)]^2_i \sum_{i=1}^{\mathcal{N}}[\vec{\mathcal{T}}^{(K)}) ]^2_i}}  \\
\end{pmatrix}.
\end{equation}

Since the normalized $L^2$ distance between two close images is close to zero, we take the inverse of this value \footnote{The reader should note that the planted attractors are asymptotically stable, which means that an infinite amount of time is required to reach the attractor exactly. In this situation, the difference between the target and the output image of our Neural Mass Network Model, at finite time, will be numerically very small. By taking the inverse of the normalized square difference, we obtain an indicator for the argument to consider.}. In other words, we use the inverse of $(\vec{m}^{(j)}_f)$. By normalizing the new vector $(\vec{m}^{(j)}_f)^{-1}$ to $1$ and applying the $\arg\,\max$ on it, we can define the accuracy ($\psi$) as:
\begin{equation} \label{magnetization}
    \psi=\frac{1}{|\mathcal{D}_{test}|}\sum_{j=1}^{|\mathcal{D}_{test}|} \delta(\arg\max \big( \vec{q}^{(j)} \big) ),
\end{equation}
where $|\mathcal{D}_{test}|$ is the cardinality of the test set,  $[\vec{q}^{(j)}]_{l}=\frac{[(\vec{m}^{(j)}_f)^{-1}]_{l}}{\sum_{i=1}^{K}[(\vec{m}^{(j)}_f)^{-1}]_i}$ and 

\begin{equation}
   \delta(\arg\max \big( \vec{q}^{(j)} \big))= \begin{cases} 1 & \text{if } \arg\max \big( \vec{q}^{(j)} \big) = \mathcal{T}'^{(j)}    \\ 0 & \text{otherwise} \end{cases}.
\end{equation}

$\delta(\arg\max \big( \vec{q}^{(j)} \big))$ is, therefore, a binary random variable that allows us to estimate the probability of success of our algorithm. 

The  computational complexity of our model, after the training process is complete, is quantified as $O(\mathcal{N}^2)$, resulting from simple matrix-vector multiplication during the integration of the dynamics, where \(\mathcal{N}\) represents the size of the input.

\section{Experimental Validations}\label{sec::results}

This section presents the results obtained from our analysis. To facilitate understanding, we divide the analysis into two subsections:
\begin{itemize}
\item \textit{Training from Scratch}: The first subsection examines the performance of classifying images on the MNIST and Fashion MNIST test sets when the Neural Mass Network Model is trained from scratch. For details on these datasets, see \ref{F-MNIST}.

\item \textit{Model Fine-Tuning with Pre-Trained Weights}: The second subsection explores the test set performance of a Wilson-Cowan model for metapopulation in combination with  convolutional neural networks \citep{Li2021CNN} or a Pre-Trained Transformer \citep{vaswani2017attention}. In the first case we will analyse the performance of the combination across MNIST, Fashion-MNIST, CIFAR10, and TF-FLOWERS. While in the second case, we perform a classification for the sentiment analysis, i.e. IMDB dataset \citep{zm1y-b270-20}. For details on these datasets, see \ref{App1}.
\end{itemize}

\subsection{Training from Scratch}

\begin{figure*}
\centering
\includegraphics[scale=0.35]{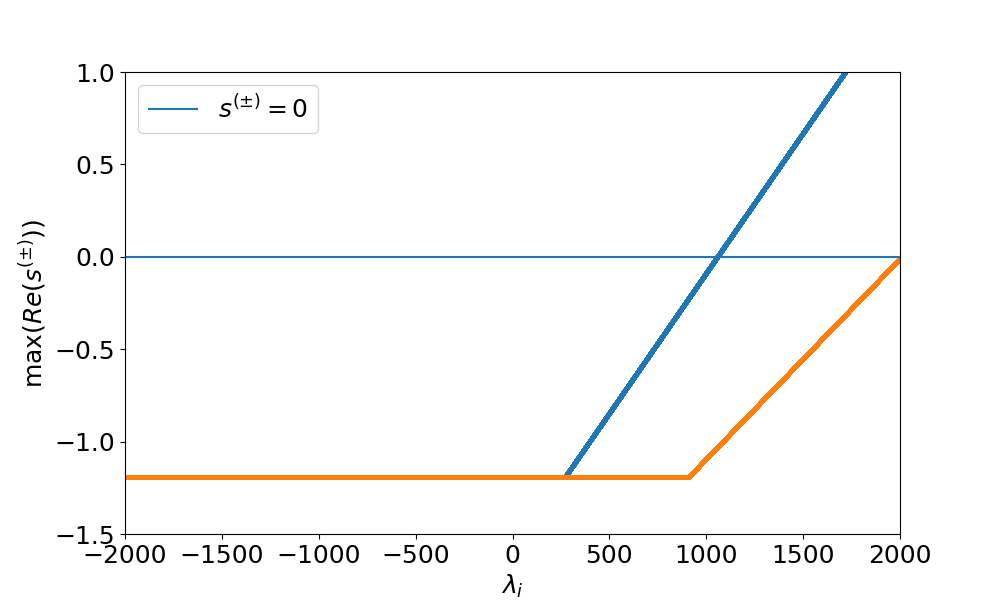}
\includegraphics[scale=0.35]{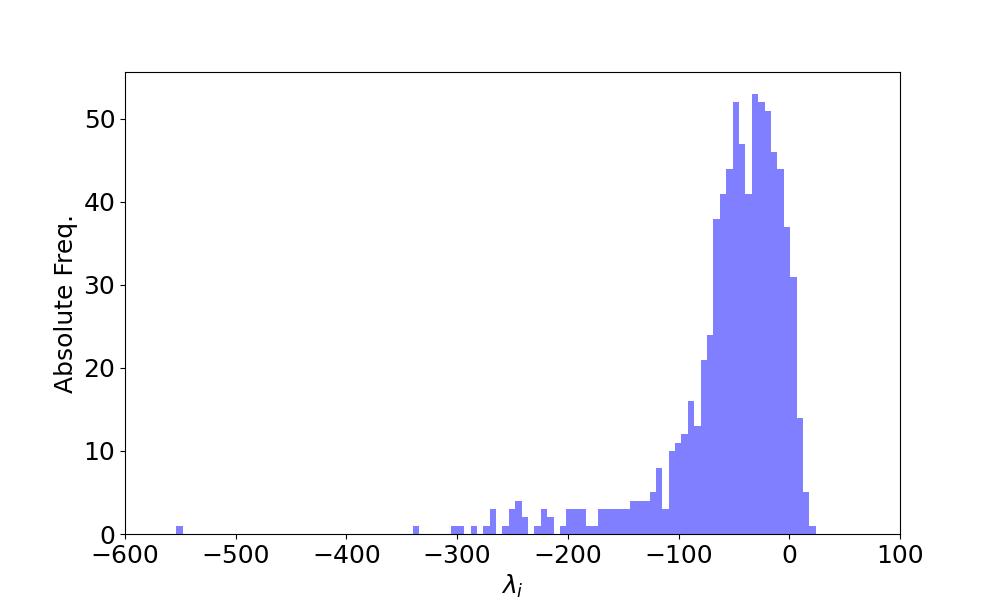}
\caption{\textbf{Stability after training.} \textbf{Top panel}: The figure shows the maximum value of the real part of the eigenvalues $s^{(\pm)}$ as a function of the parameter $\lambda_i$. The two curves, with different colors, identify respectively the two different steady states. \textbf{Bottom panel}: The figure shows the histogram of eigenvalues of matrix $\mathbf{A}$, trained on MNIST dataset. All of them satisfy the stability conditions.
The parameters here employed read $\omega_{II}=1$, $\omega_{IE}=0.0$, $\omega_{EI}=2$, $\omega_{EE}=7.2$,
$\alpha_E=1.5$, $\alpha_I=0.4$, $h_E=-1.2$, $h_I=0.1$, $\gamma=0.85$, $\beta_E=3.7$, $f^{(1)}_{E}=0.25$, $f^{(2)}_{E}=0.65$, $\beta_I=1$, $f^{(1)}_{I}=0.5$, $f^{(2)}_{I}=0.5$, $\mathcal{N}=784$.}
\label{fig_stability_final}
\end{figure*}

Our initial analysis focuses on the stability of the system for classifying images of the MNIST and Fashion MNIST datasets. According to the stability requirements established in Section \ref{sec::linearstab}, the eigenvalues of the Jacobian matrix must have negative real parts to ensure asymptotic stability at the embedded attractors. Top panel of Fig. \ref{fig_stability_final} visually displays the region where the eigenvalues of matrix $\mathbf{A}$ allow to satisfy this property. 

A critical distinction between the top panel of Fig. \ref{fig_stability_final} and Fig. \ref{fig_stability} lies in the value of the parameter $\gamma$, which plays a key role in governing the time scales within the standard Wilson-Cowan model. While traditionally it is set to $0.25$, our work explores the impact of optimizing $\gamma$ during the training process. In Fig. \ref{fig_stability}, $\gamma$ begins with its typical value of $0.25$. As stated before, a population in each vertex is composed by $80\%$ of excitatory and $20\%$ inhibitory neurons. However, during training, it undergoes optimization and converges to an optimal value of $0.839 \pm 0.031$ for the MNIST dataset ( for Fashion MNIST results see Tab. \ref{tableresultmnistandfashiongamma}). In this case, we have that a population in a vertex is composed by $54\%$ of excitatory and $46\%$ inhibitory neurons. These values suggest that, for our particular Neural Mass Network Model, a mixed population 
offers significant advantages for classification. However, we are working with a coarse-grained model of metapopulation comprising excitatory and inhibitory subpopulations. As a result, discrepancies with a real brain may occur. 
In fact, besides the discrepancy in the ratio between inhibitory and excitatory subpopulations at a single vertex, the topological structure of the learned graph also does not follow assortative modularity, where nodes connect densely within their own community and sparsely to nodes outside their community. Instead, we obtain a fully connected weighted graph.  In contrast to this topological structure, we observe that our adjecency matrix $\mathbf{A}$ remains asymmetric throughout all the analyses performed in this manuscript. Indeed, asymmetries in predicted communication efficiency reflect neurobiological concepts of functional hierarchy and correlate with directionality in resting-state effective connectivity, as analyzed using spectral dynamic causal modeling \citep{patankar2020path}.  
Moreover, a principle of brain organization is that reciprocal
connections between cortical areas are functionally asymmetric \citep{FRASSLE2021117491}.

\begin{table}[h!]
\centering
\begin{tabular}{|c|c|}
\hline
 & $\gamma (\sigma_{\gamma})$ \\ 
\hline
MNIST & 0.839(31) \\ 
\hline
Fashion MNIST & 0.922(35) \\ 
\hline
\end{tabular}
\caption{Results of $\gamma$  over five different training runs for the MNIST and Fashion MNIST datasets ($\mathcal{N}=784$). The numbers in parentheses represent the standard deviation of the mean and they refer to the last digits. Number of epochs for each single realization was set to $525$, with batch-size equal to $200$ for MNIST, while was set to $350$, with batch-size equal to $200$, for FASHION-MNIST.}
\label{tableresultmnistandfashiongamma}
\end{table}

The bottom panel of Fig. \ref{fig_stability_final}  illustrates the distribution of the final eigenvalues of $\mathbf{A}$ for the MNIST dataset. As evident from the figure, all eigenvalues reside within the designated region satisfying the imposed stability constraint. By combining the findings from both panels of Fig. \ref{fig_stability_final}, we conclude that the investigated system adheres to the stability requirements established in Section \ref{sec::linearstab}.  

\begin{figure*}
\centering
\includegraphics[scale=0.35]{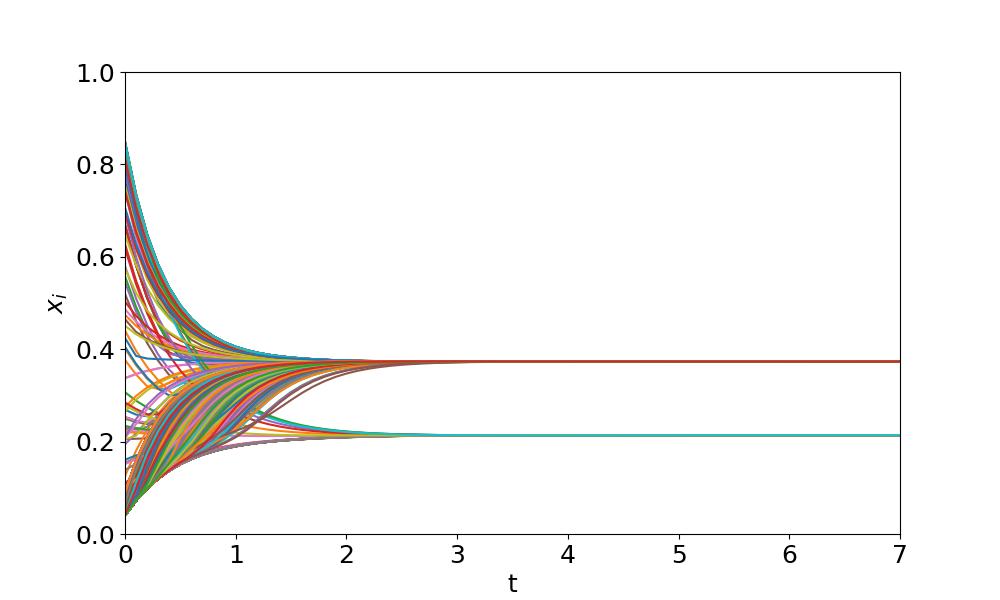}
\includegraphics[scale=0.35]{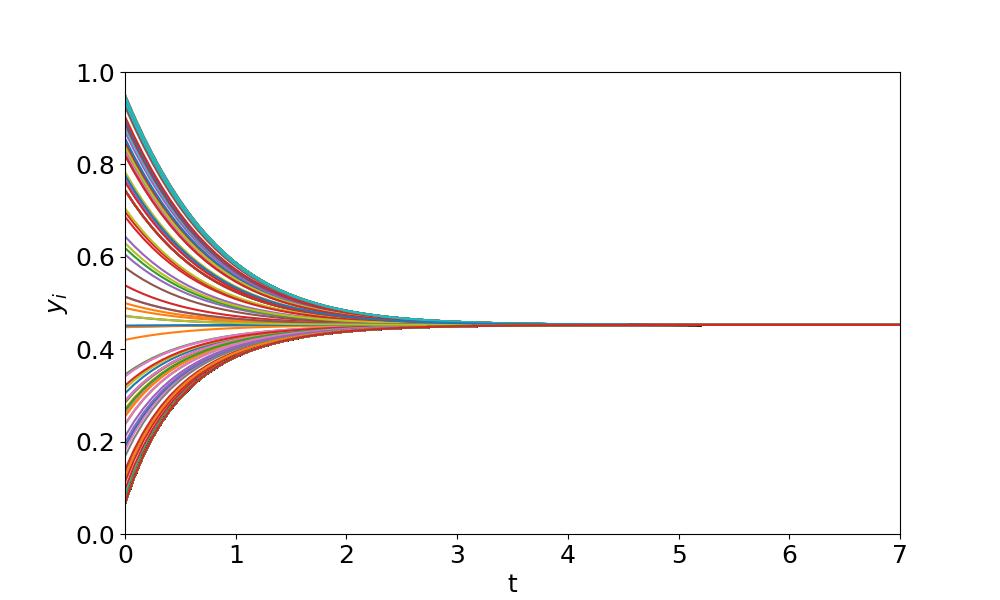}
\caption{\textbf{Dynamics after training.} \textbf{Top panel}: Evolution of all  pixels from an MNIST image under the excitatory species dynamics. \textbf{Bottom panel}: Evolution of the same pixels from the same MNIST image under the inhibitory species dynamics.
The parameters here employed read $\omega_{II}=1$, $\omega_{IE}=0.0$, $\omega_{EI}=2$, $\omega_{EE}=7.2$,
$\alpha_E=1.5$, $\alpha_I=0.4$, $h_E=-1.2$, $h_I=0.1$, $\gamma=0.85$, $\beta_E=3.7$, $f^{(1)}_{E}=0.25$, $f^{(2)}_{E}=0.65$, $\beta_I=1$, $f^{(1)}_{I}=0.5$, $f^{(2)}_{I}=0.5$, $\mathcal{N}=784$.}
\label{fig_dynamics}
\end{figure*}

Having analyzed the distribution of eigenvalues and confirmed the stability of the attractors in our Wilson-Cowan model for metapopulation, we can now delve into how this stability affects the model's dynamic behavior. Fig. \ref{fig_dynamics} presents visualizations of this connection.

This figure, containing two panels, visualizes the temporal evolution of the degrees of freedom for both excitatory and inhibitory species in our system. As previously established, our model is governed by a system of continuous ordinary differential equations in both time and space, i.e., equation \eqref{eq_stoc_vec}. The top and bottom panels depict the dynamics of all $784$ degrees of freedom associated with the excitatory and inhibitory species, respectively. Notably, the model rapidly converges to its steady state within a short timeframe of just $t = 2.5$. While the inhibitory species exhibits slightly slower dynamics (bottom panel), this delay does not significantly impact the convergence of the excitatory species towards their steady state. In fact, after $t = 1.5$ units, the excitatory population's dynamics become fully dominated by the stable attractors, indicating their rapid approach to a steady state.

\begin{figure*}
\centering
\includegraphics[scale=0.25]{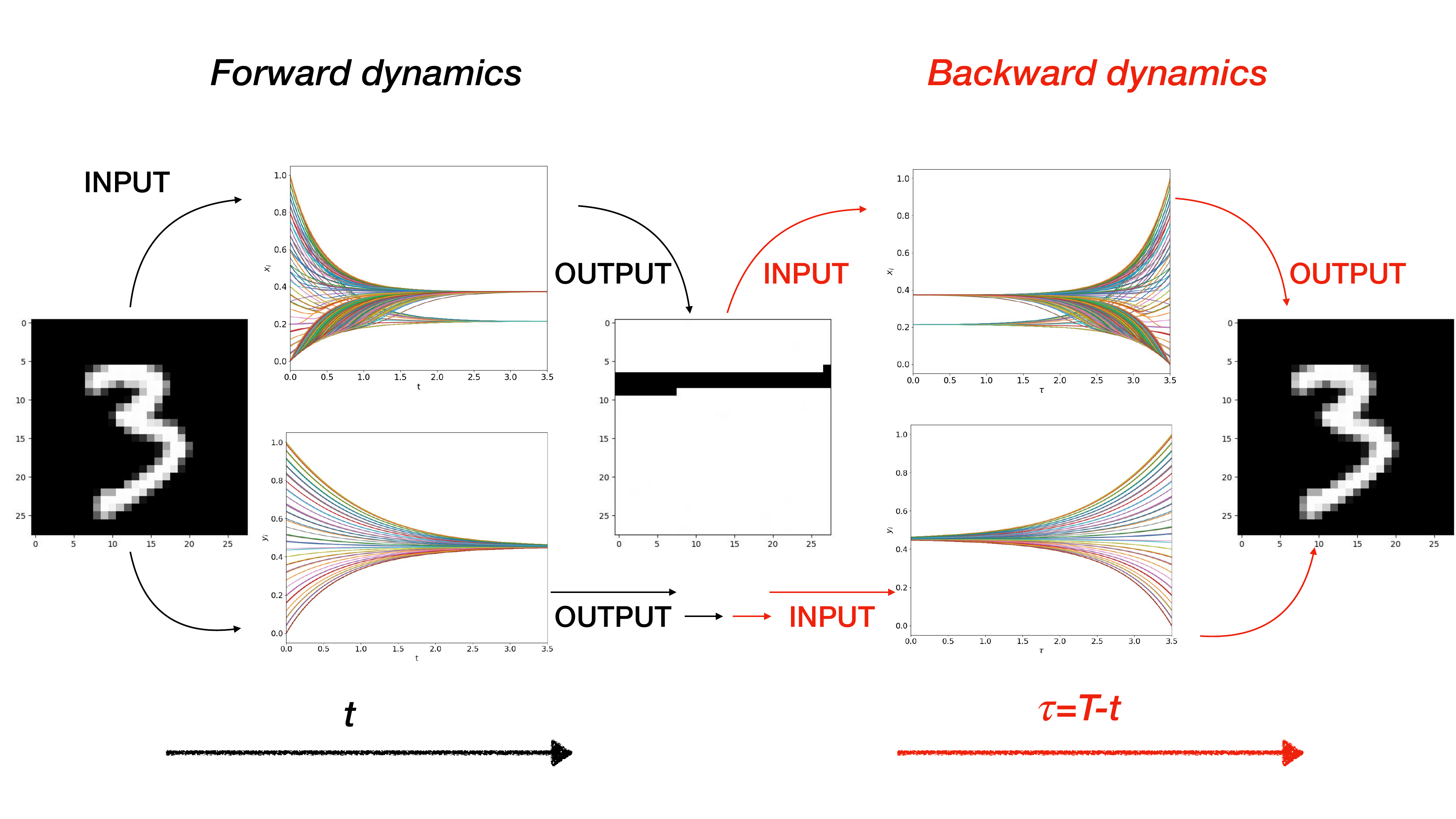}
\caption{\textbf{Cartoon of invertible property.} This figure illustrates the forward and backward dynamics of our Wilson-Cowan model for metapopulation through a cartoon representation. The algorithm takes an image as input, accurately classifies it, and subsequently reconstructs the original image entirely based on the final state of the forward dynamics. 
The parameters here employed read $\omega_{II}=1$, $\omega_{IE}=0.0$, $\omega_{EI}=2$, $\omega_{EE}=7.2$,
$\alpha_E=1.5$, $\alpha_I=0.4$, $h_E=-1.2$, $h_I=0.1$, $\gamma=0.85$, $\beta_E=3.7$, $f^{(1)}_{E}=0.25$, $f^{(2)}_{E}=0.65$, $\beta_I=1$, $f^{(1)}_{I}=0.5$, $f^{(2)}_{I}=0.5$, $T=3.5$, $\Delta t=0.0001$, $\mathcal{N}=784$.}
\label{fig_invertibility}
\end{figure*}

At this point, we can assess the classification performance of our biologically inspired learning algorithm on the MNIST dataset. Utilizing the metric defined in equation \eqref{magnetization}, we achieve, on averaged, an accuracy of $\psi=98.13\%$ (with the best performance at $\psi_{best}=98.16\%$). Comparing to a multilayer perceptron (MLP) (with one hidden $\tanh$ layer with $764$ neurons, one output softmax layer with $10$ neurons, and with a cross entropy loss function), our algorithm exhibits an accuracy only marginally lower ($0.13\%$ difference against the best performance). All the results, also for the Fashion MNIST dataset, are presented in Tab. \ref{tableresultmnistandfashion}. 

\begin{table}[h!]
\centering
\begin{tabular}{|c|c|c|}
\hline
  & $\psi (\sigma_{\psi})$ & $\psi^{MLP} (\sigma_{\psi^{MLP}})$\\ 
\hline
MNIST &   0.9813(3) &  0.9829(13)\\ 
\hline
Fashion MNIST &  0.8839(19) & 0.8955(30)\\ 
\hline
\end{tabular}
\caption{Results of  accuracy of our model (first column) and  accuracy of MLP (second column), over five different training runs for the MNIST  and Fashion MNIST datasets ($\mathcal{N}=784$). The numbers in parentheses represent the standard deviation of the mean and they refer to the last digits. Number of epochs for each single realization was set to $525$, with batch-size equal to $200$ for MNIST, while was set to $350$, with batch-size equal to $200$, for FASHION-MNIST.}
\label{tableresultmnistandfashion}
\end{table}

To compare our method with other approaches in the current literature, we reviewed studies on the MNIST and Fashion-MNIST datasets that utilized various techniques with attractors in their dynamics, as in Hopfield networks that are Recurrent Neural Networks and biological inspired. However, these methods, at best, achieved performance metrics below $70\%$ for the MNIST dataset\cite{Belyaev_2020, leonelli2021effective}, and $63\%$  for the Fashion-MNIST dataset \cite{fachechi2022outperforming}. Even with a supervised approach that transforms Hebb's rule into a genuine learning rule, the accuracy only reached $94\%$ for MNIST and $84\%$ for Fashion-MNIST \cite{alemanno2023supervised}.

Additionally, we compared our results with those obtained by Krotov and Hopfield in their work \cite{Krotov2016}. In this study, the authors demonstrated that the Hopfield network can be modified by using a non-linear (e.g., polynomial) function in the Hamiltonian, allowing it to store a polynomial number of memories, rather than just a linear one. These memories are stable attractors of the discrete dynamics. Furthermore, the authors showed that these modified networks, known as Dense Associative Memory (DAM) networks, can also function as classifiers when a continuous non-linear function is applied in the asynchronous dynamics and a stochastic gradient descent is used to optimize the weights.

For instance, DAMs achieve a test error of $1.6 \%$ on the MNIST dataset, while our Wilson-Cowan model for metapopulation achieves a test error of $1.9 \%$. In other words, our model performs on par with DAMs.

This model also offers a simple invertible property. By employing a straightforward transformation, $\tau=T-t$ \citep{ marino2023solving}, which effectively reverses the dynamics of the system, we can integrate the dynamics backwards, starting from the terminal position of the forward dynamics, and rebuild the initial condition from which we started with. The modified evolutionary law for this backward integration becomes $\dot{\mathbf{z}}=-G(\mathbf{z})$, with the temporal derivative now taken concerning the variable $\tau$. A simple cartoon is presented in Fig. \ref{fig_invertibility}. 

Our analysis was performed on simple grayscale image datasets of size $28 \times 28$. Such an analysis is computationally fast, as showed in   \ref{ScalingAnalysis} for different image sizes. We performed also ablation experiments on our model, and such experiments are presented in \ref{ablationexp}. However, testing colored images, such as those in the CIFAR-10 and TF-FLOWERS datasets, which have sizes $32 \times 32 \times 3$ and $224 \times 224 \times 3$ respectively, can be time-consuming due to the increased dimensionality of these images. To avoid such bottlenecks, we test our model with CNNs, which are biologically inspired. Additionally, we investigate whether the well-known concept of \textit{transfer learning} \citep{Hospedales2021transferlearning} can be applied to our model. These analyses are presented in the following subsection.

\subsection{Model Fine-Tuning with Pre-Trained Weights}
In this section, we present experimental validation of our Neural Mass Network Model in combination with CNNs and Transformers.

\subsubsection{ Wilson-Cowan model for metapopulation in combination with a Convolutional Neural Network}

\begin{figure*}
\centering
\includegraphics[scale=0.22]{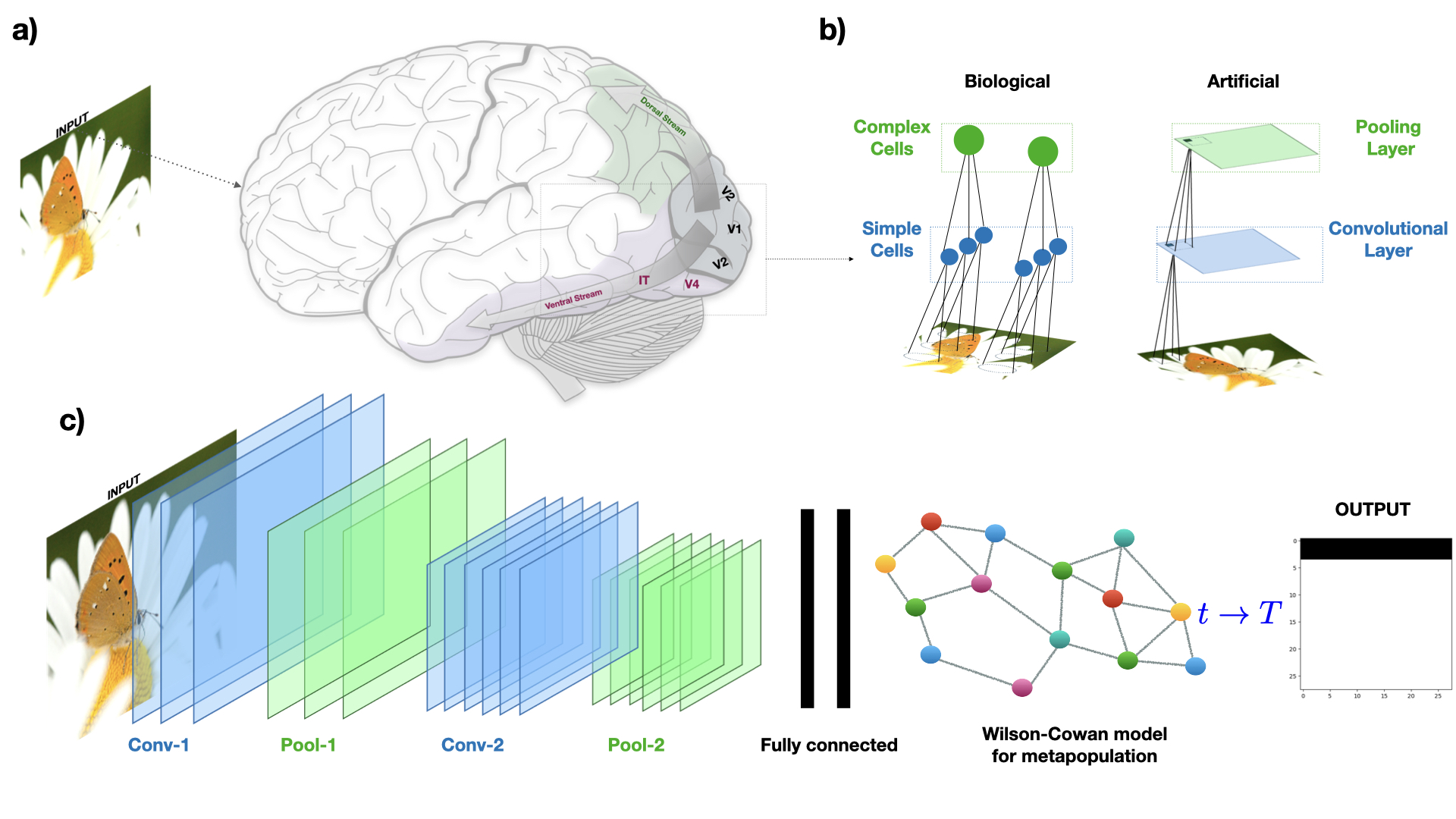}
\caption{\textbf{Description of biological and artificial object recognition.} The figure illustrates a schematic representation of how object recognition and classification, in the brain and artificially, should work. Panel \textbf{a)} simply describes the position of the visual cortex (see \ref{VCortex}).  Panel \textbf{b)} shows the relationship between components of the visual system (\textbf{Biological}) and the base operations of a CNN (\textbf{Artificial}). \textbf{Biological}: cartoon of simple cells (blue) and complex cells (green) \citep{hubel1959receptive}. Simple cells have preferred location in the image (dashed ovals). Complex cells receive input from many simple cells and thus have more spatially invariant responses. \textbf{Artificial}:  the first convolutional layer (blue) is produced by applying a small filter (square box)  to every location in the image. Such operation creates a collection of feature maps. For each 
feature map, for example, the maximal activation in the square blue box can be taken. Such an operation downsamples the image and leads to a complex cell-like plane (green). Panel \textbf{c)} illustrates a sketch of the CNN model that we use. It has a series of convolutional and pooling layers. It terminates with a set of dense layers that are connected to a Wilson-Cowan model for metapopulation ($\mathcal{N}=784$). Given an input, the CNN feeds forward through all its layers to arrive at our model, which iteratively converges to the planted stationary state.}
\label{CNNfig}
\end{figure*}

Computational models validate intuitions about how a system works by providing a way to test those intuitions directly. They offer a means to explore new hypotheses in an ideal experiment. CNNs are an example. They came from pioneering research in neuroscience, which gave insights into the nature of visual processing in mammals including humans. 
In such a pioneering research,  \citep{hubel1959receptive} discovered two major cell types in the primary visual cortex (V1) of cats (see Fig. \ref{CNNfig}, Panel \textbf{b)}). The first type, the simple cells, respond to bars of light or dark when placed at specific spatial locations. The second type, complex cells, have less strict response profiles. These complex cells are likely receiving input from several simple cells, all with the same preferred orientation but with slightly different preferred locations. 
From such observations, Fukushima developed the Neocognitron \citep{fukushima1980neocognitron}, a precursor to modern CNN. This computational model contains two main cell types: the S-cells and the C-cells. The S-cells are named after simple cells and replicate their basic features on a plane, as well as the C-cells (named after complex cells) that are nonlinear function of several S-cells coming from the same plane but at different locations. After a plane of simple and complex cells representing the basic computations of V1, the Neocognitron simply repeats the process again. 
From these ideas, finally CNNs arose.  
A CNN is a type of deep learning model \citep{Li2021CNN} specifically designed for processing structured grid data, like images. CNNs automatically and adaptively learn spatial hierarchies of features from input images through convolutional layers, pooling layers, and fully connected layers. For more on architectures and applications of CNNs, we suggest the reviews \citep{Lindsay2021, zhao2024review}.

CNNs can mimic and simulate the representation of visual information along the ventral stream (see Fig. \ref{CNNfig}, Panel \textbf{a)}). In particular, the activity of the artificial units of CNNs predicts the activity of real neurons in animals, with very high accuracy. For example, in \citep{Yamins2014performance} the authors showed the existence of a strong correlation between a biologically plausible hierarchical neural network model’s categorization performance and its ability to predict individual inferior temporal (IT) neural unit response data. Furthermore, the activity of units from the last layer of the network best predicted IT activity and the penultimate layer best predicted V4, in visual cortex. Many other works have been developed in this direction, demonstrating that several deep CNN architectures exhibit similar performance to human and monkey object classification \citep{Lindsay2021, Rajalingham7255}.

Merging, therefore, a CNN with our Wilson-Cowan model for metapopulation seems reasonable. Indeed, both models appear to be biologically inspired and, when validated separately, address some aspects of brain function.

Such a new model could, in principle, retain all the validated aspects of a visual system provided by the CNN part and then, at the end of the ventral cortical visual system (i.e., the IT), incorporate a recurrent network—our Wilson-Cowan model for metapopulation—that contains some long-term memories, which we define as our planted attractors. The IT region, indeed, is a brain region where visual perception meets memory and imagery \citep{Miyashita1993}.  Long-term memory refers to the brain's process of taking information from short-term memory and creating long-lasting memories \citep{she2024temporal}.
These memories are stable and easily accessible, like our planted attractors. Moreover, long-term memories can include information related to activities learned through practice and repetition, similar to how a learning algorithm operates. However, validating such a computational model is extremely challenging. We must begin, therefore, with the initial steps of implementing the model and analyzing its performance on various visual classification tasks. We defer in-depth analyses for validating the model with neurobiological data to future publications.

We begin by analyzing the MNIST and Fashion MNIST datasets to determine if adding a CNN in front of our Wilson-Cowan model for metapopulation can improve classification accuracy. The architectures and training hyperparameters used are detailed in \ref{F-MNIST}. The results are presented in Table \ref{tableresultcnn}.

\begin{table}[h!]
\centering
\begin{threeparttable}
\begin{tabular}{|c|c|c|}
\hline
  & $\psi (\sigma_{\psi})$ & SOTA\\ 
\hline
MNIST &  0.9931(8) &  0.9987\tnote{a}\\ 
\hline
Fashion MNIST & 0.9135(16) & 0.9691\tnote{b}\\ 
\hline
CIFAR10 & 0.8659(21) & 0.9950\tnote{c} \\ 
\hline
TF-FLOWERS & 0.8485(37) & 0.98\tnote{d}\\ 
\hline
\end{tabular}
    \begin{tablenotes}
      \footnotesize \tiny
      \item[a] \citep{BYERLY2021545}
      \item[b] \citep{tanveer2021fine}
      \item[c] SOTA obtained by using transformer architecture \citep{dosovitskiy2021an}. The best VGG-16 \citep{SimonyanZ14a} has an accuracy of 0.93\citep{giuste2020cifar}.
      \item[d] SOTA obtained by using \href{https://blog.tensorflow.org/2020/05/bigtransfer-bit-state-of-art-transfer-learning-computer-vision.html}{transformer architecture}\citep{bit2020}.
    \end{tablenotes}
\end{threeparttable}
\caption{Results of accuracy of our model (first column) and state-of-the-art (SOTA) (second column), over five different training runs for the MNIST, Fashion MNIST, CIFAR10 and TF-FLOWERS datasets. The numbers in parentheses represent the standard deviation of the mean and they refer to the last digits.}
\label{tableresultcnn}
\end{table}

As the reader can see, the results obtained by our simple CNN combined with our Wilson-Cowan model for metapopulation are satisfactory. Indeed, for the MNIST dataset, our new model seems to be close to the state-of-the-art, differing by only $0.5\%$, on average. However, for the Fashion MNIST dataset, the discrepancy increases to $5\%$, on average, compared to the state-of-the-art.

For both datasets, the $\gamma$ parameter appears to remain close to the accepted values for a biological model (see Tab. \ref{tableresultcnngamma}). Indeed, the populations of excitatory neurons in all cortical areas should be around $70-80\%$ \citep{DEFELIPE1992563}, so our optimal values fell within the right range for the ratio between inhibitory and excitatory neurons.

\begin{table}[h!]
\centering
\begin{tabular}{|c|c|}
\hline
 & $\gamma (\sigma_{\gamma})$\\ 
\hline
MNIST & 0.330(34)  \\ 
\hline
Fashion MNIST & 0.252(8) \\ 
\hline
CIFAR10 & 0.247(1)\\ 
\hline
TF-FLOWERS & 0.25(0) \\ 
\hline
\end{tabular}
\caption{Results of $\gamma$  over five different training runs for the MNIST, Fashion MNIST, CIFAR10 and TF-FLOWERS datasets. The numbers in parentheses represent the standard deviation of the mean and they refer to the last digits.}
\label{tableresultcnngamma}
\end{table}

We then decided to implement a VGG-16 architecture (see \ref{vgg16app}) for classifying the CIFAR-10 dataset. In this case, we observed that the accuracy obtained is on average $\psi=86.59\%$ (with the best performance at $\psi_{best}=86.83\%$), thanks to the deep convolutional architecture used. Although our model lacks specific dropout and batch normalization layers, and data augmentation techniques that are typically used in the best VGG-16 implementations, our performance remains reasonably high compared to many other CNN architectures. For further comparison, the reader shall refer to the performance benchmarks available at this \href{https://paperswithcode.com/sota/image-classification-on-cifar-10}{link} \footnote{ \href{https://paperswithcode.com/sota/image-classification-on-cifar-10}{https://paperswithcode.com/sota/image-classification-on-cifar-10} }.

For the last analysis, we chose to check if transfer learning still works with our model. In general, transfer learning can be used to improve performance on a task A for which training data is in short supply by using data from a related task B, for which data is more plentiful. The two tasks should have the same kind of inputs, and there should be some commonality between the tasks so that low-level features, or internal representations, learned from task B will be useful for task A. When data for task A is very scarce, we might simply retrain the final layer of the network. In contrast, if there are more data points, it is feasible to retrain several layers. This process of learning parameters using one task that are then applied to one or more other tasks is called pre-training \citep{bishop2023deep}.

\begin{figure*}
\centering
\includegraphics[scale=0.25]{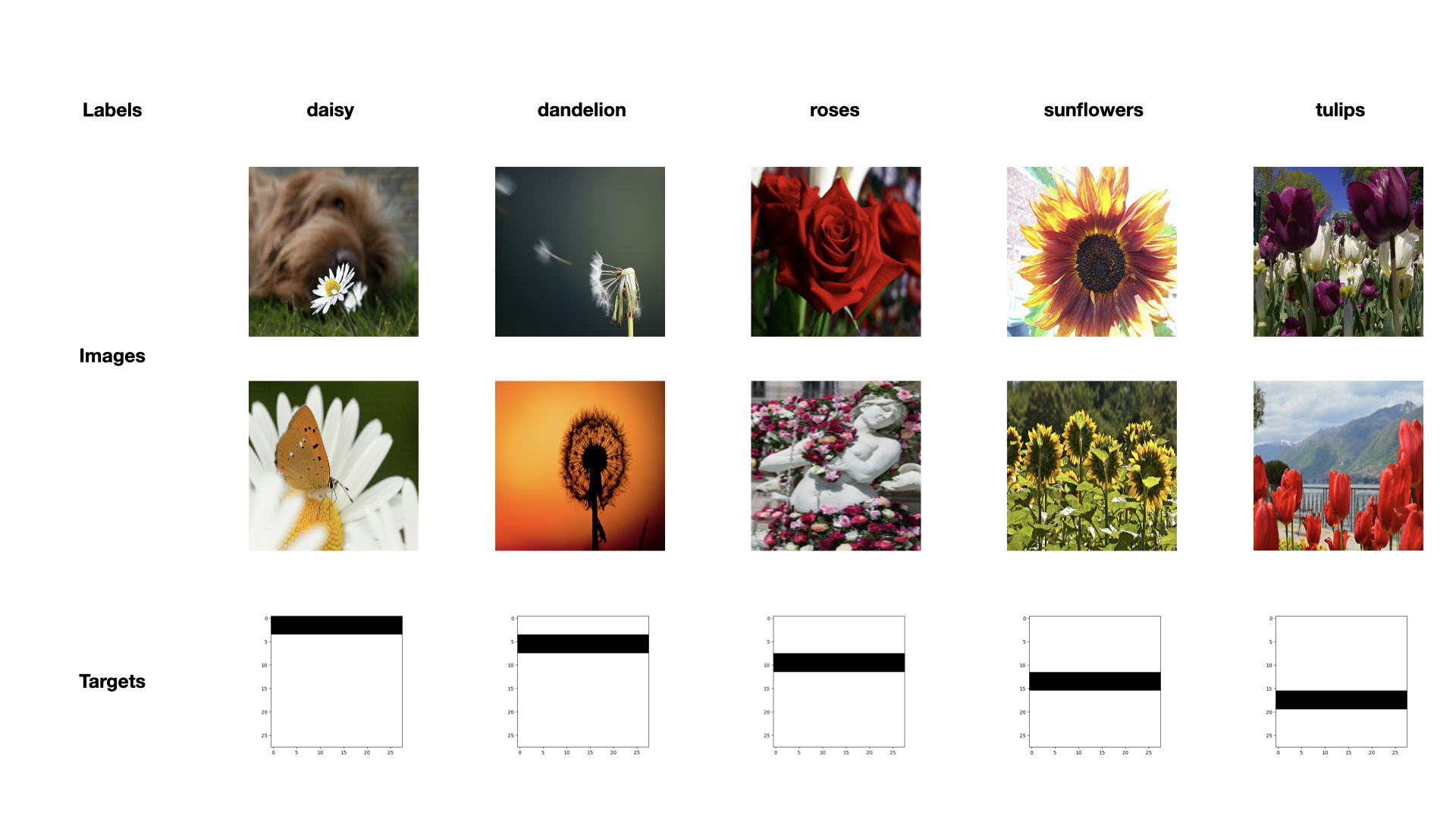}
\caption{\textbf{Description of TF-FLOWERS dataset.} The figure illustrates the composition of TF-FLOWERS dataset. The first row shows the labels of the examples found in the second and the third row. The fourth row presents the five target images we created to serve as attractors for the dynamics ($\mathcal{N}=784$).}
\label{figflowers}
\end{figure*}

We, therefore, confined ourselves to work with a small dataset, TF-FLOWERS (see ref{tfflowers} for details). This dataset is composed of $3670$ colored $224 \times 224 \times 3$ images of flowers, categorized into $K=5$ distinct classes: daisy, dandelion, roses, sunflowers, and tulips. An example of such images is presented in Fig. \ref{figflowers}. It is our task A, and as the reader can see, the dataset is scarce in the number of images, but not in the details of each image. We, therefore, decided to use the VGG16 model trained on the IMAGENET dataset\footnote{The IMAGENET  dataset \citep{imagenetdataset2009} comprises $14$ million natural images each of which has been hand labelled into one of nearly $22000$ categories. A subset of images comprising $1000$ non-overlapping categories is often used for pre-training models in deep learning. The fact to have so many categories made the problem much more challenging because, if the classes were distributed uniformly, random guessing would have an error rate of $99.9\%$.} (task B) for transfer learning, and then train just the last layers, as well as our Wilson-Cowan model for metapopulation (see  \ref{tfflowers}).

The performance of this algorithm, as described in Table \ref{tableresultcnn}, reaches on average $84.85\%$ (with the best performance at $\psi_{best}=85.28\%$). This demonstrates that transfer learning is effective, as expected, when combining a pre-trained CNN with our Wilson-Cowan model for metapopulation. However, the high number of parameters reduces the biological plausibility of the model. Indeed, the parameter $\gamma$ can be considered a free parameter that does not significantly affect the algorithm's performance, if chosen reasonably.

\subsubsection{ Wilson-Cowan model for metapopulation in combination with a Pre-Trained Transformer}

We have previously observed that the use of CNNs significantly enhances the accuracy of models. These networks draw inspiration from biological systems and exhibit remarkable similarities to specific regions of the human brain. Recently, CNNs have also been employed to describe and understand how emotional patterns are integrated into the human visual system \citep{Kragel2019}. However, CNNs fall short in replicating another fundamental capability of the human brain: the comprehension and production of language.

Human language processing is one of the most fascinating capabilities of our brain. It relies on a set of interconnected brain areas in the frontal and temporal lobes, typically in the left hemisphere, forming a network. This language network supports both comprehension (spoken, written, and signed) \citep{Deniz7722} and production \citep{Hu2022precision}. It has been extensively studied to understand its sensitivity to linguistic regularities at multiple levels \citep{regev2024high}.

Given the structural similarities between our Wilson-Cowan model for metapopulation and the topological configuration of a language network, it seems reasonable to apply our model to this context. We aim to determine if the Wilson-Cowan model could effectively perform a simple classification task within this framework. To investigate this, we combined our approach with a novel machine learning architecture: transformers.

Transformers, as described by  \citep{vaswani2017attention} in their seminal work, have become the most trending topic in natural language processing (NLP) due to their outstanding performance in capturing formal linguistic competence—i.e., the knowledge of rules and statistical regularities of language. However, they exhibit limitations in functional linguistic competence, which involves the practical use of language in real-world situations \citep{MAHOWALD2024517}. 

This section explores the potential synergy between our model and a transformer architecture. We explore this synergy to address the limitations of the Wilson-Cowan model for metapopulation in extrapolating the complex structural patterns embedded in language. Therefore, we utilize a transformer architecture, specifically BERT \citep{devlin2018bert}, to extract the most important features from the text for a classification task with our Neural Mass Network Model.

To achieve this, we chose to perform sentiment analysis on the IMDB dataset (see \ref{imdb}). Sentiment analysis is an NLP technique used to determine the emotional tone behind a body of text. This process involves analyzing text data to identify and categorize the sentiments expressed, typically as positive or negative. The IMDB dataset consists of movie reviews from the Internet Movie Database, with each review labeled as either positive or negative. Therefore, the classification task is a straightforward binary classification.

To be precise, we fine-tune  a BERT model in combination with our Wilson-Cowan model (see \ref{imdb}), which in this case, as a reminder, has two stable attractors planted in the dynamics. Given the simplicity of our binary classification task, we slightly modify the training process. We use binary cross-entropy as the loss function. To achieve this, we transform the output of the Wilson-Cowan model for metapopulation into a probability of belonging to a class. This is done by taking the normalized inverse vector, obtained by normalizing the $L^2$ distance between the terminal condition of the system of Wilson-Cowan equations and the respective target, i.e. equation \eqref{magnetization}.

\begin{table}[h!]
\centering
\begin{tabular}{|c|c|c|c|}
\hline
 & $ \psi (\sigma_{\psi})$ & $\psi_{BERT} (\sigma_{\psi_{BERT}})$ & SOTA\\ 
\hline
IMDB & 0.8746(22)  &  0.8830(3) &  0.9668 \citep{csanady2024llambert} \\ 
\hline
\end{tabular}
\caption{Results of accuracy ($\psi$) of our model (first column), accuracy of BERT without our Wilson-Cowan model and trained with the same hyper parameters of our model (second column), state-of-the-art (third column), over five different training runs for the IMDB. The numbers in parentheses represent the standard deviation of the mean and they refer to the last digits.}
\label{tableresultBERT}
\end{table}

The accuracy results are presented in Tab. \ref{tableresultBERT}. As the reader can see, the performance of our model in synergy with BERT is comparable to that of BERT alone. However, in comparison to the state-of-the-art (SOTA) models, our performance is $9\%$ points lower. This discrepancy is due to the fact that current SOTA classifications are dominated by massive transformers, which have parameter counts that are two or more orders of magnitude greater than those of the BERT model. As detailed in \ref{App2}, our choice of architecture was also influenced by the hardware limitations we faced.

\section{Conclusion}\label{sec::conclusion}

In this manuscript, we have presented a Wilson-Cowan model for metapopulation capable of learning to classify images and text. We began by defining the model and detailing the methodology for embedding stable attractors within the metapopulation dynamics. Subsequently, we explained how to train this model using a supervised learning framework. We then conducted various numerical analyses to demonstrate the high accuracy this model can achieve across different classification tasks.

Although our method, even when combined with other computational neural models, achieves high accuracy, it does not surpass the state-of-the-art deep learning algorithms for classification tasks. This gap is primarily due to our choice of architecture, which, while once state-of-the-art, has been eclipsed by models with significantly more learning parameters, exceeding our hardware capabilities. However, as demonstrated, our model's performance is still close, but not equal, to the maximum accuracy achieved by those advanced models. There are two main reasons for this discrepancy: (i) We embed and enforce stability on our attractors (targets), thus limiting the solution space within which the learning algorithm can search for an optimal solution. (ii) We did not employ any image preprocessing techniques, such as data augmentation, or advanced engineering tricks in building the architecture, as our focus was not on achieving state-of-the-art performance but on demonstrating the functionality of our biologically inspired model.

A careful reader may have wondered why we have not claimed that our model is a plausible biological model but have only described it as biologically inspired. This distinction arises primarily because, despite numerous connections to biological behavior and topological similarities with brain structures, our model is trained using the backpropagation algorithm. This training method prevents us from identifying our model as a truly plausible biological model.

However, backpropagation can be viewed as an efficient way to achieve reasonable parameter estimates, which can then be subjected to further testing. Even if backpropagation is considered merely a technical solution, the trained model may still serve as a good approximation of neural systems. Currently, many researchers are exploring new supervised learning optimization algorithms that are more biologically valid, i.e., neurobiologically plausible methods by which the brain could adjust its internal parameters to optimize objective functions \citep{song2024inferring}. Future publications will aim to validate new biologically plausible Wilson-Cowan models for metapopulation that can effectively learn visual or textual patterns.

\section*{Data availability statement}
The numerical codes used in this study and the data that support the findings are available at \href{https://github.com/RaffaeleMarino/Learning_in_WilsonCowan}{github.com/RaffaeleMarino/Learning\_in\_WilsonCowan}.

\section*{Acknowledgments}

R.M. and L.C. are supported by \#NEXTGENERATIONEU (NGEU) and funded by the Ministry of University and Research (MUR), National Recovery and Resilience Plan (NRRP), project MNESYS (PE0000006) "A Multiscale integrated approach to the study of the nervous system in health and disease" (DR. 1553 11.10.2022).
F.D.P. thanks Gruppo Nazionale di Fisica Matematica of Istituto Nazionale di Alta Matematica for partial financial support. F.D.P is supported by Next GenerationEU PRIN 2022 research project “The Mathematics and Mechanics of nonlinear wave propagation in solids” (grant n°2022P5R22A).

\section*{Competing Interests Statement}
The authors have no competing interests to declare

\appendix
\section{Datasets, CNN Architectures, hyper-parameters and training tricks}\label{App1}
All the architectures in this manuscript were chosen to fit with our hardware (see \ref{App2}), ensuring that results could be obtained within a reasonable time frame.

\subsection{MNIST and Fashion MNIST}\label{F-MNIST}

The MNIST dataset \citep{deng2012mnist} is composed by $70000$ grey scale handwritten images of size $28\times 28$ in $K=10$ classes. The dataset is divided in $60000$ images for training set and $10000$ for test set.

The Fashion-MNIST dataset \citep{xiao2017fashion}  is composed by $70000$ grey scale Zalando's article images of size $28\times 28$ in $K=10$ classes. The dataset is divided in $60000$ images for training set and $10000$ for test set.

The CNN for these analyses is composed by two convolutional layers with $32$ feature channels with kernel size $3\times3$, each utilizing the ReLU activation function. Each convolutional layer is followed by a max pooling layer with pool size $2\times 2$. After these layers, a flatten layer is used to convert the output for the dense layers. The CNN output is then passed to three distinct dense layers with $2048$, $1024$, and $784$ neurons, all using the ReLU activation function. The second and third dense layers are preceded by a batch normalization layer. The output of the third dense layer is then passed to our Wilson-Cowan model for metapopulation, with $\mathcal{N} = 784$. The total number of parameters is $5179521$, divided into $615441$ parameters for our Wilson-Cowan model for metapopulation and $4564080$ for the CNN part plus the three dense layers.

We used a learning rate of $0.0001$ for this analysis. We performed an initial training procedure only on the CNN part plus the three dense layers, using a mini-batch size of $10$ for $35$ epochs. Then, we conducted a complete training procedure for the entire model, including our Wilson-Cowan model for metapopulation, with a mini-batch size of $200$ for $70$ epochs and a value of $T=3.5 \Delta t^{-1}$, where $\Delta t = 0.1$. The loss function was set to be always the one in equation \eqref{eq::lossmatrix}.
    
\subsection{VGG-16}\label{vgg16app}

The VGG-16 \citep{SimonyanZ14a} model, where VGG stands for the Visual Geometry Group, who developed the model, and 16 refers to the number of learnable layers in the model, has some simple designed principles leading to a relative uniform architecture that minimizes the number of hyperparamete choices that need to be made. In principle, it was developed to take an input image having $224 \times 224 \times 3$ colored pixels (RGB channels), followed by sets of convolutional and pooling layers for downsampling. It can be also applied to smaller color images $32 \times 32 \times 3$. Here we present the original architecture as described in \citep{bishop2023deep}. 

On each convolutional layer is applied a filter of size $3\times3$ with a stride of $1$, same padding, and a ReLU activation function. Each pooling layer, instead, applies a maximum pooling operation with stride $2$, filter size $2\times2$, downsampling the number of units by a factor $4$. To be precise, the first learnable layer is a convolutional layer in which each unit  takes input from a $3\times3\times3$ tensor from the stack of input channels, and so has $28$ parameters including the bias. These parameters are shared across all units in the feature map for that channel. There are $64$ such feature channels in the first layer, giving an output tensor of $224 \times 224 \times 64$. The second layer is also convolutional and again it has 64 channels. This is followed by the first maximum pooling layer that gives feature maps of size $112 \times 112$. The third and the fourth layer are again  convolutional, of dimensionality $112 \times 122$ with $128$ channels. This is again followed by a maximum pooling layer to give a feature map size $56 \times 56$, followed by three convolutional layers with $256$ channels and followed again by another maximum pooling layer to give a feature map size $28 \times 28$. The output of this layer is feed forwarded to another set of three convolutional layers each having $512$ channels, followed by another maximum pooling layer, which downsamples to feature maps of size $14 \times 14$. This is followed by three more convolutional layers, with $512$ channels, and another maximum pooling layer for downsampling to $7 \times 7$, with $512$ channels. Finally, for making classification, three dense layers are added. In this manuscript, the last three dense layers are modified (see main text, and later subsections).

\subsection{CIFAR10} \label{cifar}
The CIFAR-10 dataset \citep{cifar} is composed by $60000$ colored images of size $32\times 32 \times 3$ in $K=10$ classes (airplane, automobile, bird, cat, deer, dog, frog, horse, ship, and truck), with $6000$ images per class. There are $50000$ training images and $10000$ test images.
The CNN for this analysis is composed by the VGG16 architecture (see \ref{vgg16app}). After these layers, a flatten layer is used to convert the output for the dense layers. The CNN output is then passed to three distinct dense layers with $2048$, $1024$, and $784$ neurons, all using the ReLU activation function. The second and third dense layers are preceded by a batch normalization layer. The output of the third dense layer is then passed to our Wilson-Cowan model for metapopulation, with $\mathcal{N} = 784$. The total number of parameters is $19294817$, divided into $615441$ parameters for our Wilson-Cowan model for metapopulation and $18679376$ for the CNN part plus the three dense layers.

We used a learning rate of $0.0001$ for this analysis. We performed an initial training procedure only on the CNN part plus the three dense layers, using a mini-batch size of $10$ for $70$ epochs. Then, we conducted a complete training procedure for the entire model, including our Wilson-Cowan model for metapopulation, with a mini-batch size of $200$ for $70$ epochs and a value of $T=3.5 \Delta t^{-1}$, where $\Delta t = 0.1$. The loss function was set to be always the one in equation \eqref{eq::lossmatrix}.

\subsection{TF-FLOWERS}\label{tfflowers}
The TF-FLOWERS dataset \citep{Xu2022} is composed by $3670$ colored $224\times 224 \times 3$ images of flowers, categorized into $K=5$ distinct classes:  daisy, dandelion, roses, sunflowers, and tulips. We chose $90\%$ of the dataset as training images and $10\%$ test images.
The CNN for this analysis is composed by the VGG16 architecture (see \ref{vgg16app}), with pre-trained weights given by IMAGENET dataset. After these layers, a flatten layer is used to convert the output for the dense layers. The CNN output is then passed to four distinct dense layers with $4096$, $2048$, $1024$, and $784$ neurons, all using the ReLU activation function. The second and third dense layers are preceded by a batch normalization layer. The output of the third dense layer is then passed to our Wilson-Cowan model for metapopulation, with $\mathcal{N} = 784$. The total number of parameters is $129460641$, divided into $615441$ parameters for our Wilson-Cowan model for metapopulation and $128845200$ for the CNN part plus the three dense layers.

We used a learning rate of $0.001$ for this analysis. We performed an initial training procedure only on the CNN part plus the three dense layers, using a mini-batch size of $10$ for $70$ epochs. Then, we conducted a complete training procedure for the entire model, including our Wilson-Cowan model for metapopulation, with a mini-batch size of $32$ for $100$ epochs and a value of $T=3.5 \Delta t^{-1}$, where $\Delta t = 0.1$. The loss function was set to be always the one in equation \eqref{eq::lossmatrix}.

\subsection{IMDB}\label{imdb}

The IMDB (Internet Movie Database) dataset \citep{zm1y-b270-20} is a comprehensive and widely-used dataset in the field of machine learning and data analysis, particularly for tasks involving natural language processing (NLP) and sentiment analysis. This dataset contains extensive information on movies, television shows, and other forms of visual entertainment. The labeled data set consists of $50000$ IMDB movie reviews, specially selected for sentiment analysis. The sentiment of reviews is binary, meaning the IMDB rating $< 5$ results in a sentiment score of $0$, and rating $\geq7$ have a sentiment score of $1$. No individual movie has more than $30$ reviews. The $25000$ review labeled training set does not include any of the same movies as the $25000$ review test set. In addition, there are another $50000$ IMDB reviews provided without any rating labels.

For this analysis we consider transformer language model based on encoders, which are models that take sequences as input and produce fixed length vectors, such a class labels, as output. More precisely, we use the Bidirectional Encoder Representations from Transformers (BERT) architecture, which is a pre-trained language model \citep{devlin2018bert}. Unlike other language representation models, BERT is designed to pretrain deep bidirectional representations from unlabeled text by jointly conditioning on both left and right context in all layers.  For our analysis we used the so called \textbf{BERT$_{\text{BASE}}$} model \citep{devlin2018bert}. It is composed by $12$ transformer layers, with hidden sizes equal to $768$ and with $12$ self-attention heads. The total number of parameters for BERT is set to $109482241$.

This language model is then associated with a dropout layer with a parameter $p = 0.5$ and a dense layer of $512$ neurons with a sigmoid activation function. Following this, we integrate our Wilson-Cowan model for metapopulation (with $\mathcal{N}=512$), ending up with a final model with $110138626$ . As stated in the main text, we perform a modification for this particular classification task. Specifically, we do not compare the final state of the dynamics with the respective planted eigenvector anymore. Instead, we apply a non-linear transformation to the final state, normalizing the inverse of equation \eqref{magnetization}, to obtain a straightforward output for our $K = 2$ class problem. Consequently, we derive a probability to be in one of the two classes, allowing us to apply binary cross-entropy as the loss function and utilize binary accuracy to test the performance of our learning algorithm. We used a learning rate of $0.00003$, and we fixed  the number of epochs for the fine-tuning of the whole network at $10$, with mini-batch size equal to $4$. The final time $T$ was set to be equal $4.0 \Delta t^{-1}$, with $\Delta t=0.1$. For the tokenization of the dataset, we have followed the tutorial given at this \href{https://www.tensorflow.org/text/tutorials/classify_text_with_bert}{link} \footnote{ \href{https://www.tensorflow.org/text/tutorials/classify_text_with_bert}{https://www.tensorflow.org/text/tutorials/classify\_text\_with\_bert}} .

\section{Hardware Specification}\label{App2}
All the analyses presented in this manuscript were run on a Lenovo 256GB RAM  workstation with 2 GPU NVIDIA-RTX A5500, 24GB RAM each. 

\section{Visual Cortex}\label{VCortex}
The visual cortex (grey, purple and green) is the primary cortical region of the brain that receives, integrates, and processes visual information relayed from the retinas. It is in the occipital lobe of the primary cerebral cortex, which is in the most posterior region of the brain. The visual cortex divides into five different areas (V1 to V5) based on function and structure. In figure \ref{CNNfig} Panel \textbf{a)}, only the positions of V1, V2, and V4 are presented for the sake of simplicity. Inferior Temporal (IT) is the cerebral cortex on the inferior convexity of the temporal lobe in primates including humans. It is crucial for visual object recognition and is considered to be the final stage in the ventral cortical visual system (grey and purple). The ventral stream transforms visual inputs into perceptual representations that embody the enduring characteristics of objects and their spatial relations. The ventral stream begins with V1, goes through visual area V2, then through visual area V4, and to the inferior temporal cortex. The ventral stream, is associated with form recognition, object representation and storage of long-term memory \citep{Schneider1969, MILNER2008774}.  The dorsal stream (grey and green) begins with V1, goes through area V2, then to the dorsomedial area and middle temporal area and to the posterior parietal cortex. The dorsal stream’s job is to mediate the visual control of skilled actions, such as reaching and grasping, directed at objects in the world. To do this, the dorsal stream needs to register visual information about the goal object on a moment-to-moment basis, transforming this information into the appropriate coordinates for the effector being used \citep{Schneider1969, MILNER2008774}.

\section{Scaling Analysis}\label{ScalingAnalysis}

In this appendix, we present an analysis of the scaling behavior \cite{Marino2023} of our neural mass network model when applied to classification tasks on the MNIST and Fashion MNIST datasets. We analyze how accuracy scales with the image size. To achieve this, we create new datasets from the original ones, where the image sizes are $14 \times 14$, $17 \times 17$, $21 \times 21$, $24 \times 24$, $28 \times 28$, $31 \times 31$, and $35 \times 35$.

Table \ref{tabMNISTscaling} shows the accuracy as a function of the input size $\mathcal{N}$ for MNIST, while Table \ref{tabFMNISTscaling} for Fashion MNIST. For both tables,  The first column identifies the size of the image, the second the value of the $\gamma$ parameter, the third one the accuracy of the model, and the last column is the time for a single epoch in seconds on our hardware \ref{App2}.

\begin{table}[ht]
\centering
\begin{tabular}{|c|c|c|c|}
\hline
$\mathcal{N}$  & $\gamma$ & $\psi$ & time for each epoch [s] \\
\hline
196   & 0.86046 & 0.9746  & 2  \\
\hline
289   & 0.8822  & 0.98    & 3  \\
\hline
441  & 0.8311  & 0.9818  & 4  \\
\hline
576  & 1.066   & 0.9815  & 5  \\
\hline
784  & 0.839   & 0.9813  & 7  \\
\hline
961  & 0.827   & 0.9781  & 10 \\
\hline
1225& 1.044   & 0.981   & 12 \\
\hline
\end{tabular}
\caption{Scaling of neural mass network model on varying image sizes for MNIST dataset. Total number of epochs 525, batch size 200.}
\label{tabMNISTscaling}
\end{table}

\begin{table}[ht]
\centering
\begin{tabular}{|c|c|c|c|}
\hline
$\mathcal{N}$  & $\gamma$ & $\psi$ & time for each epoch [s] \\
\hline
196   & 0.8625 & 0.8752  & 2  \\
\hline
289   & 1.064  & 0.8794    & 3  \\
\hline
441  & 0.971  & 0.8788  & 4  \\
\hline
576  & 0.910   & 0.8853  & 5  \\
\hline
784  & 0.992  & 0.8839  & 7  \\
\hline
961  & 1.016   & 0.8847  & 10 \\
\hline
1225& 1.016   & 0.887   & 12 \\
\hline
\end{tabular}
\caption{Scaling of neural mass network model on varying image sizes for Fashion MNIST dataset. Total number of epochs 350, batch size 200.}
\label{tabFMNISTscaling}
\end{table}

\newpage
\section{Ablation experiments }\label{ablationexp}

In this appendix, we present several ablation experiments \cite{meyes2019ablation} on our neural mass network model. We focus on the MNIST dataset and test the hypotheses required to enable the model to function as a classifier.

We begin by removing the stability criterion from our model while keeping all other conditions fixed. Specifically, we force the eigenvalues of the matrix $\Lambda$ to fall outside the stability region. However, when we do this, the optimization process becomes infeasible, preventing us from transforming the Wilson-Cowan model for metapopulation into a classifier.

From this reason, stability is always preserved in our model. Next, we performed ablation on the planted eigenvectors. In this setup, we maintained the stability regions for the trainable eigenvalues, retained the target structure using the fixed points of the dynamics (as presented in this work), kept the zero eigenvalues fixed, and removed the planted eigenvectors, allowing the model to learn them automatically. We observed that the model was able to construct these eigenvectors; however, it achieved a lower accuracy, classifying correctly only $97.83\%$ of the test set images after training.

We then analyzed the model under the condition that the fixed eigenvectors are not required to remain within the kernel of $\mathbf{A}$, while all other components of the model were kept unchanged. In this experiment, we set the fixed eigenvalues to $0.1$. Under these conditions, the accuracy decreased to $97.28\%$, compared to the case in which the fixed eigenvectors remain in the kernel of $\mathbf{A}$.

In conclusion, we observed that the combination of all methods used to construct the Wilson-Cowan model for metapopulation as a classifier is essential for achieving high accuracy. Furthermore, we found that the stability criterion is the most crucial component of the model; without it, our model cannot be used for classification tasks

\bibliographystyle{apa} 
\bibliography{apssamp}

\end{document}